%% file: qA.tex
\documentclass[aps,preprint,noshowpacs,nofootinbib,superscriptaddress,groupedaddress]{revtex4}  
\usepackage{graphicx} 
\usepackage{dcolumn}   
\usepackage{bm}        
\usepackage{amssymb}   
\usepackage{amsmath}

\newcommand{\eg}{\eq}
\input{defs}

\begin{document}

\begin{flushright}
CERN-TH-2019-201\\
HIP-2019-37/TH\\
INT-PUB-19\\
\end{flushright}

\title{Gluon Radiation from a classical point particle II: dense gluon fields }

\author{K. Kajantie}
\affiliation{Helsinki Institute of Physics, FI-00014 University of Helsinki, Finland}
\email{keijo.kajantie@helsinki.fi}
\author{Larry D. McLerran}
\affiliation{Institute for Nuclear Theory, University of Washington, Box 351550, Seattle, WA, 98195, USA}
\email{mclerran@me.com}
\author{Risto Paatelainen}
\affiliation{Theoretical Physics Department, CERN, CH-1211 Gen\`eve 23, Switzerland}
\email{risto.sakari.paatelainen@cern.ch}
\affiliation{Helsinki Institute of Physics, FI-00014 University of Helsinki, Finland}

\date{\today}

\begin{abstract}
The goal of this paper is to extend the results of Ref.~\cite{Kajantie:2019hft}, where formulae were derived for 
gluonic radiation for a high energy nucleus colliding with a classical colored particle. In Ref.~\cite{Kajantie:2019hft}, we computed the amplitudes for radiation in the fragmentation region
of the particle for a dilute gluonic field. In this paper, we compute the radiation by solving the fluctuation equations of the dense background field
in a specific gauge which makes it simple to solve the asymptotic radiation from an initial condition immediately after the passage of
the nucleus. We identify and compute two components of gluon radiation, a bulk component which extends to the central region and bremsstrahlung, which
may give rise to an experimentally observable intensity peak in the target fragmentation region.

\end{abstract}


\maketitle

\section{Introduction}
\label{sec:intro}
\input{corrintro}

\section{Transverse radiation field}
\label{eom}
\input{corrradfield}

\input{corrbbiv2}

\input{conclu}

\appendix

\newpage
\section{Green's function integrals over the classical quark currents}
\label{appA}

\input{appA}

\section{Wilson line correlators}
\label{appC}
\input{appC}

\section{Average over the classical color space}
\label{appB}
\input{appB}

\begin{acknowledgments}
RP is supported by the European Research Council, grant no. 725369 and LM was supported by 
the U.S. DOE under Grant No. DE-FG02- 00ER41132. KK and RP thank F. Gelis, T. Lappi, N. Armesto and W. Van Der Schee for discussions.
\end{acknowledgments}

\input{biblio}
\end{document}

%% file: defs.tex
\newcommand{\fig}{Fig.~}

\newcommand{\eq}{Eq.~}

\newcommand{\eqs}{Eqs.~}
\newcommand{\reff}{Ref.~}

\newcommand{\yt}{\mathbf{y}}
\newcommand{\xt}{\mathbf{x}}
\newcommand{\zt}{\mathbf{z}}
\newcommand{\bt}{\mathbf{b}}
\newcommand{\kt}{\mathbf{k}}
\newcommand{\qt}{\mathbf{q}}
\newcommand{\rt}{\mathbf{r}}
\newcommand{\pt}{\mathbf{p}}
\newcommand{\vt}{\mathbf{v}}
\newcommand{\ot}{{\mathbf{0}}}
\newcommand{\hht}{\mathbf{h}}
\newcommand{\tr}{{\mathrm{Tr}}}

\newcommand{\uw}{{\mathcal{U}}}

\newcommand{\be}{\begin{equation}}
\newcommand{\ee}{\end{equation}}
\newcommand{\ba}{\begin{eqnarray}}
\newcommand{\ea}{\end{eqnarray}}
\newcommand{\nn}{\nonumber \\}
\newcommand{\nr}[1]{(\ref{#1})}
\newcommand{\rmi}[1]{{\mbox{\scriptsize #1}}}

\newcommand{\fra}[2]{{\textstyle{\frac{#1}{#2}\,}}}  

%% file: corrintro.tex
In our first paper, Ref. \cite{Kajantie:2019hft}, we considered a particle being scattered by a high energy nucleus which was treated as a sheet of colored glass \cite{McLerran:1993ni,McLerran:1993ka}.  The particle (which we call a quark) being scattered is treated in an approximation where its color field is very weak compared to that of the high energy nucleus, in analogy with the treatment of Refs. \cite{Kovchegov:1998bi,Dumitru:2001ux}. In these papers, and many others (see e.g. Refs. \cite{Kovner:1995ja} - \cite{mt3}), gluonic radiation in the central region of the scattering is computed. In this paper, we compute for the first time the distribution of gluonic radiation emitted by this scattering process in the fragmentation region of the target quark. The scattering process is represented by a classical current containing a static target quark and a scattered quark with given momentum. We shall identify a component which depends on the nuclear saturation scale $Q_s^A$ and is independent of the momentum of the scattered quark. However, in this classical current approximation there is another component, bremsstrahlung, which depends on the scattered quark momentum, is independent of nuclear effects and gives wrong large transverse momentum $k_T$ behavior. This issue arises due to lack of quantum recoil in the classical current so that to give correctly the gluon distribution for bremsstrahlung contribution at larger $k_T$ the model should be improved. The most interesting region of computation, however, is when the $k_T \ll Q_s^A$. In this region, the majority of particles are produced and our computation within a classical current approximation should give a valid description of particle production in the fragmentation region of the target quark.

Investigating this problem is  interesting since it solves an old problem: the initial conditions for matter produced in the fragmentation region of high energy scattering \cite{Anishetty:1980zp, km}.  In early computations, it was found that the matter produced in the fragmentation region is baryon rich.  In the collision the baryons are compressed and there is gluon radiation produced that heats up the system.  These early arguments provided a justification for the assertion that one produced a quark gluon plasma in heavy ion collisions.  One purpose of this paper is to update these early computations in the modern context of the theory of the Color Glass Condensate (CGC) \cite{Iancu:2000hn,Ferreiro:2001qy}.

In later work, we intend to use these initial conditions to  provide a theoretically based formulation of the early time evolution of the baryon rich glasma produced in such collisions \cite{Kovner:1995ja,Kovner:1995ts,McLerran:2018avb}.  Perhaps eventually such consideration may be useful for hydrodynamics studies of the fragmentation region of high energy collisions \cite{Denicol:2018wdp}, and the insight engendered by such consideration may allow a more solid theoretical treatment of heavy ion collisions at energies below which one makes a baryon free central region.

\begin{figure}[!t]
\begin{center}
\includegraphics[width=0.65\textwidth]{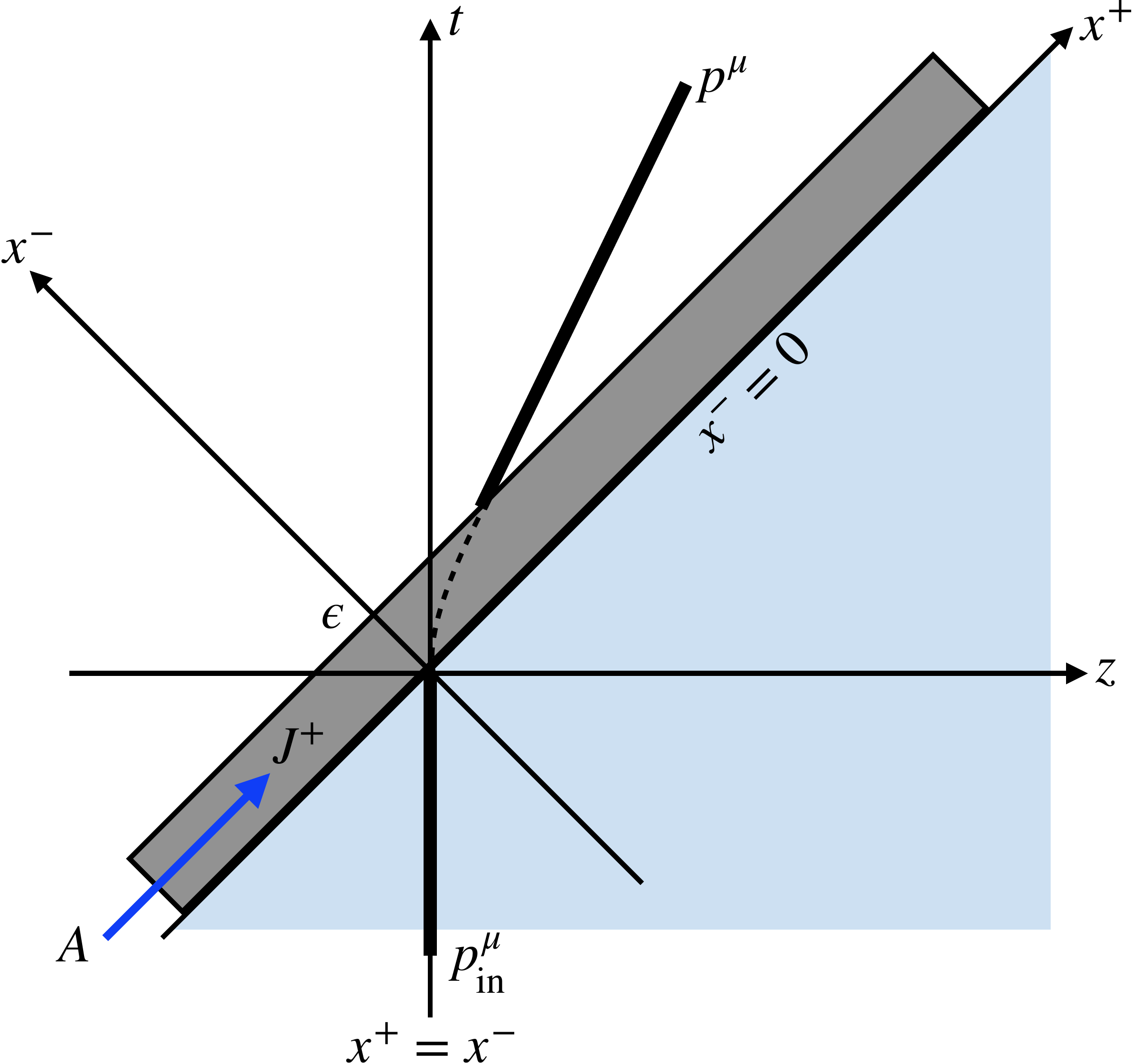}
\end{center}
\caption{\small Nucleus $A$ moving along the light-cone $x^-=0$ collides with a static quark in the origin,
interacts with its color electric field and accelerates it to a constant momentum (thick lines). The nuclear vacuum equivalent background color field in LC gauge $A^-=0$ is chosen so that it is nonzero
at $x^-<0$ before the collision (light blue area) and zero after it. The asymptotic color fluctuation field, which gives the gluon distribution,
can then be simply solved from initial conditions formulated at $x^-=\epsilon\to0$. }
\label{collision}
\end{figure}

A simple physical interpretation of what we do is as follows. We first study the process (see \fig\ref{collision})
\be
A+q(p_\rmi{in})\to X+g(k)
\ee
in the target fragmentation region of the initial quark so that initial 4-momentum
\footnote{We use mostly plus light-cone coordinates,  $v\equiv v^\mu=(v^+,v^-,\vt)$, $|\vt|=v_T$ ,  $v^{\pm}=-v_{\mp}$ and $v^i=v_i$. Bold face quantities are two-dimensional transverse vectors, $\vt=(v_1,v_2)=(v_i)$, $dv_1dv_2=d^2v$. The scalar product of two four-vector is $\xt \cdot \yt = -x^+y^- - x^-y^+ + \xt \cdot \yt$. The rapidities $y,\,y_p$ always
refer to the rest frame of the target particle, $y=y_\rmi{CM}+y_\rmi{beam}$.} 
is $p_\rmi{in}=(\fra{m}{\sqrt2},\fra{m}{\sqrt2},{\bf 0})$. The strongly contracted nucleus $A$ is moving along the light-cone (LC)
$x^-=0$ and its large color field \cite{McLerran:1993ni,McLerran:1993ka} interacts with the color electric field of the quark. We want to compute the distribution $dN/(dyd^2k)$ of produced gluons with momentum $k=(k^+,k^-,\kt)$, $k^\pm=\fra{k_T}{\sqrt2}e^{\pm y}$, $k^2=-2k^+k^-+k_T^2=0$. Interactions of the nuclear sheet with this field will be shown (see \ref{subbulk}) to be the main source of
gluon production in the fragmentation region of the static quark. 
To do this computation, we need a dense nuclear background field
$A^\mu$  and how it interacts with a small fluctuation field $a^{\mu}$, which gives rise to the electric field of the initial
static quark. This interaction follows from  Yang-Mills fluctuation equations. The produced gluon distribution is then computed by finding 
$a^{\mu}$ at asymptotic infinity. We shall call this the bulk distribution and it will be the main quantitative contribution to the goal of this work.

A crucial technical detail of this work is  that the nuclear background gauge is chosen so that 
it only populates the region of space $x^-<0$ before the nucleus. Usually it is chosen so that it is excited
after the passage of the nucleus. With this choice the fluctuation field for $x^->0$ is a free field and solving
it simply requires computing the initial condition at $x^-=0^+$. Much of the later work is devoted to finding this initial
condition.

In addition to the gluons the final state also contains the scattered quark of momentum 
$p=(p^+,p^-,\pt)$, $p^\pm=\fra{m_T}{\sqrt2}e^{\pm y_p}$, $p^2=-2p^+p^-+p_T^2=-m^2$ and we shall also consider 
more differentially the process
\be
A+q(p_\rmi{in})\to X+q(p)+g(k)
\ee
assuming $p$ is a given constant momentum. The process then is gluon production by a classical quark current. We shall derive
the fluctuation equations for this extended case and show that it naturally splits up to the bulk contribution described above 
and a bremsstrahlung contribution. This bremsstrahlung contribution, however, lacks quark recoil and has to be improved.

Quantum mechanics enters by integrating over an ensemble of classical sources. Diagrammatically, one may describe the setup by
saying that the bulk contribution arises by emission of a gluon before the collision, which then interacts with the nucleus.
We shall show (see \ref{resdisc}), that crucial role for getting a physically valid result is also played by interference with gluons
emitted before the collision. Emission of a gluon after the collision, which does not further interact with nucleus, gives rise to bremsstrahlung. Superficially, the setup 
is quite different from the case in which both the projectile and target move along opposite light-cones. This frame is
reached by boosting $y\to y_\rmi{CM}=y-y_\rmi{beam}$, i.e., by $y\to\infty$.

The rest of the paper is structured as follows. The classical current is defined and discussed in Sect. \ref{subcurrent}.  
The field equations describing the interactions between the current of the initial and accelerated quark and the nuclear background
field are described in Section \ref{subeom} and solved in Section \ref{subsol}. The main result is presented after
introducing quantum effects in Section \ref{subquant} and discussed in Section \ref{resdisc}. 
Bulk and bremsstrahlung contributions and their interference are studied in Sections \ref{subbulk}, \ref{bremint} and \ref{interference}. Finally, in Section \ref{totaldist} we show full numerical results for the gluon distribution in the target fragmentation region. Appendices \ref{appA}, \ref{appC} and \ref{appB}  contain details of several subproblems.

%% file: corrradfield.tex
The physics of our problem contains  (see \fig\ref{collision}) a nucleus moving along the light-cone which collides with 
an initially static quark and accelerates it. The goal is
to compute the distribution of gluons produced in this process. In our model of QCD, the nucleus is represented by a strong background color field, the quark by a classical quark current and produced gluons by a transverse radiation field. In this section, we shall define the quark current, determine
equations of motion for a small fluctuation field $a^{\mu}$ and solve them for the radiation field. Part of this material is in paper I, see \reff\cite{Kajantie:2019hft}.

\subsection{The quark current}\label{subcurrent}
Let us first consider the gauge field $A^{\mu}$ induced by the dense projectile nucleus in the longitudinal light-cone gauge $A^-=0$, where the gauge field is entirely $A^+ = A^+(x^-,\xt)$. The current $J^{\mu}$ for the nucleus corresponds to a Lorentz contracted sheet of infinitesimal thickness at $x^- = \epsilon$ with $\epsilon \to 0$, and it reads 
\begin{equation}
\label{currentJ}
 J^{\mu a}(x^-,\xt)= \delta^{\mu, +} \delta(x^-) \rho^a(\xt).
\end{equation}
Here the color charge density on the nuclear sheet is $\rho^a(\xt)$ and the $SU(N)$ adjoint color index $a$ runs from $1,\dots,N^2-1$.
The current is independent of $x^+$ by the extended current conservation law 
\be
D_{\mu}J^{\mu} = \partial_{\mu}J^{\mu} - ig[A_{\mu},J^{\mu}] =  \partial_+J^+ = 0. 
\ee
The classical Yang-Mills (CYM) equations with the current defined in \eq\nr{currentJ} reduce to 
\be
D_{\mu}F^{\mu\nu} = J^{\nu} \quad  \longrightarrow \quad \partial^2_T A^+(x^-,\xt) = J^+(x^-,\xt).
\ee

The dilute target is described by a single color charged particle. The classical current $j^{\mu}$ of the struck particle can be computed by using Wong's equations \cite{wong} for the trajectory $x^{\mu}(\tau)$ from the following expression
\be
j^{\mu}(x) = g \int d\tau~Q(\tau)u^{\mu}(\tau)\delta^{(4)}(x-y(\tau)),
\ee
where $u^{\mu} \equiv \dot{x}^{\mu}(\tau)$. The particle in a colored field has a classical color charge $Q(\tau) \equiv Q^a(\tau)T^a_{(F)}$, where $T^a_{(F)}$ are the generators in the fundamental representation of $SU(N)$.  The classical color charge $Q^a$ is assumed to transform as the adjoint representation of the $SU(N)$ gauge group. The average over the space of classical color charges is given by\footnote{See also the discussion in \reff\cite{Mrowczynski:2016etf}.}:
\be
\label{coloraverage}
\langle Q^aQ^b \rangle_Q = \frac{1}{2}\delta^{ab}, \quad\quad \langle Q^a \rangle_Q =0,
\ee
where the shorthand notation $\langle Q^aQ^b \rangle_Q$ is defined as
\be
\langle Q^aQ^b \rangle_Q  \equiv \int dQ\, Q^aQ^b, \quad\quad \int dQ =1
\ee
and the integration measure $dQ $ over the space of classical color charges is given in \reff\cite{Litim:2001db}.

The covariant conservation $D_{\mu}j^{\mu}=0$ implies that 
\be
\frac{dQ^a}{d\tau} -igu^{\mu}[A_{\mu},Q]^a = 0.
\ee
From this equation it follows that the color charge $Q^a$ evolves according to 
\be
Q^a \to \uw_{ab}Q^b, \quad\quad \uw(x) = \mathcal{P}\exp \left (-ig\int_{x} dy_{\mu}A^{\mu}(y)\right ),   
\ee
where $\uw$ is the path ordered adjoint Wilson line.

As shown in \fig\ref{collision}, we consider the setup in which a single quark traverses the nucleus with a constant trajectory from the backward light-cone with $u^{\mu} = p_{\rm in}^{\mu}/m$ to the forward light-cone with $u^{\mu} = p^{\mu}/m$. Since the nuclear field extends over the range $x^- \in [0, \epsilon]$  with $\epsilon \to 0$, the current $j^{\mu}$ at $x^- <  \epsilon$ is modified by color precession in the background field $A^+$.  This implies that the classical quark current for $x^- < \epsilon$ with $\epsilon \to 0$, expressed in a frame where the quark is at rest with a fixed four-momentum $p^{\mu}_{\rm in} $, can be written as  
\be
\label{eq:currentjfull1}
\begin{split}
j^{\mu a}(x^+,x^- <0,\xt) & = g \uw(\ot)_{ab}Q^b  \frac{p_{\rm in}^{\mu}}{m} \int d\tau~\delta(x^- - u^-\tau)\delta(x^+-u^+\tau)\delta^{(2)}(\xt - \mathbf{u} \tau) \\
& =  g \uw(\ot)_{ab}Q^b \mathcal{J}^{\mu}(x^+,x^- < 0 ,\xt),
\end{split}
\ee  
where we have defined
\be
\label{bigj1}
\mathcal{J}^{\mu}(x^+,x^- < 0 ,\xt) \equiv   \frac{p_{\rm in}^{\mu}}{p_{\rm in}^{-}} \delta(x^+-x^-)\delta^{(2)}(\xt).
\ee
Here the adjoint Wilson line $\uw$ for $x^- \in [0,\epsilon]$ is given by 
\be
\label{eq:Uwilsonsol}
\uw(x^-=\epsilon,\xt) = \mathcal{P}\exp \left (-ig T^a_{(A)}\int_{x^-}^{\infty} dy^- A^+_{a}(y^-,\xt) \right ),
\ee 
and $T^a_{(A)}$ are the generators of $SU(N)$ in the adjoint representation. In the limit $\epsilon \to 0$ this simplifies to  
\be 	      
\uw(\xt) = \mathcal{P}\exp \left (-ig T^a_{(A)}\Phi_a(\xt) \right ),
\ee 
where the field $\Phi_a$ satisfies $\partial^2_T \Phi_a(\xt) = \rho_a(\xt)$. After the collision at $x^- > 0$,  the quark at rest accelerates to fixed momentum $p^{\mu}$. The corresponding color current reads
\be
\label{eq:currentjfull2}
\begin{split}
j^{\mu a}(x^+,x^- > 0 ,\xt)  = gQ^a \mathcal{J}^{\mu}(x^+,x^- > 0 ,\xt),
\end{split}
\ee  
where
\be
\label{bigj2}
\mathcal{J}^{\mu}(x^+,x^- > 0 ,\xt) \equiv  \frac{p^{\mu}}{p^{-}}\delta\left (x^+ - \frac{p^+}{p^-}x^-\right )\delta^{(2)}\left (\xt - \pt \frac{x^-}{p^-}\right ).
\ee

Putting \eqs\nr{eq:currentjfull1} and \nr{eq:currentjfull2} together, we find that the final total quark current reads
\ba
j^{\mu a}(x^+,x^- ,\xt)&=& g \uw(\ot)_{ab}Q^b{p_\rmi{in}^\mu\over p_\rmi{in}^-} \delta(x^+-x^-)\delta^{(2)}(\xt)\theta(-x^-)+
\nn&&
+gQ^a \frac{p^{\mu}}{p^{-}}\delta\left (x^+ - \frac{p^+}{p^-}x^-\right )\delta^{(2)}\left (\xt - \pt \frac{x^-}{p^-}\right )\theta(x^-).
\ea

\subsection{Equations of motion}\label{subeom}

We are interested in the radiation field produced after the quark-nucleus collision; i.e. knowing the gauge fields for $x^- < 0$, what are the gauge fields far in the future at $x^->0$.  The radiation field can be obtained by adding a small perturbation $a^{\mu}$ to the background gauge field $A^{\mu}$ of the nucleus  
\be
A^{\mu} = A^{\mu} + a^{\mu} \quad\text{with}\quad J^{\mu} = J^{\mu} + j^{\mu},
\ee
and solving the corresponding CYM equations to linear order in the small fluctuation field $a^{\mu}$.  However, in the initial longitudinal light-cone gauge where the gauge field is entirely $A^+$, the gauge field is singular at $x^- = 0$. In order to avoid this unpleasant feature, it is optimal to choose a gauge where the gauge field $A^{\mu}$ is a two dimensional transverse field $A^i$ for $x^- <0$ so that it vanishes for $x^- >0$ before the passage of the nucleus.
This goal can be achieved by choosing suitably the gauge rotation that connects the longitudinal $A^+$ gauge to the transverse $A^i$ gauge. Following the discussion in \cite{Kajantie:2019hft}, this amounts to solving the equation:
\be
\partial_{-}\uw^{\dagger}(x^-,\xt) = -igA^+(x^-,\xt)\uw^{\dagger}(x^-,\xt),
\ee
where $\uw$ is the gauge rotation matrix transforming $A^+$ to zero. A solution where $\uw$ is one for $x^- >0$ is given by the adjoint Wilson line in \nr{eq:Uwilsonsol}. For $x^- <0$, the two-dimensional gauge field associated with this rotation is 
\be
-igA^i(\xt) = \uw(\xt)\partial^i \uw^{\dagger}(\xt).
\ee

In this new gauge (hereby referred as the transverse LC gauge), the equation of motion for the transverse fluctuation field $a^i$ can be written as\footnote{This is obtained from the general fluctuation equation $D^2a^{\mu} - D^{\mu}(D\cdot a) - 2igF^{\mu\nu}a_{\nu} = \tilde{j}^{\mu}$.} \cite{Kajantie:2019hft}
\be
\label{eq:flukxsmall0}
D^2a^i = \left (-2\partial^+\partial^- + D^2_T \right )a^i = \tilde{j}^{i} - D^i \frac{1}{\partial^-}\tilde{j}^{-},
\ee
where the notation $D^2_T \equiv D_kD_k$ is introduced. The current $\tilde{j}^{\mu}$  is related to the current $j^{\mu}$ in the $A^+$ gauge via the transformation  $\tilde{j}^{\mu} \to \uw j^{\mu}$. Using the relations:
\be
\begin{split}
D_kS &= \uw\partial_k(\uw^{\dagger}S)\\
D^2_TS &= \uw\partial^2_T(\uw^{\dagger}S)
\end{split}
\ee 
and noting that $[\partial^{\pm}, \uw] =0$ yields the expression
\begin{equation}
\label{eq:flukeq1}
\Box\left (\uw^{\dagger}a^i \right ) = j^i - \partial^i \frac{1}{\partial^-}j^- \equiv j^i_{\rm eff},
\end{equation}
where $\Box \equiv -2\partial^+\partial^- + \partial^2_T$ is the d'Alembert operator.

\subsection{Solving the radiation field}\label{subsol}
We will now utilize the special gauge choice made,  vanishing background field $A^i$ at $x^->0$. For $x^->0$ the gauge field then is a free field
and can simply be solved at infinity by giving the field at $x^-=0$. In \cite{Kajantie:2019hft} it was proven that the field $a_i$ is continuous
at $x^-=0$.

For $x^- <0$, the full solution of \eq\nr{eq:flukeq1} can be written down in terms of the retarded Green function:
\begin{equation}
\label{eq:solflukeq1}
a^{i}(x^+,x^- <0, \xt) = a^{i}_{\rm in} + \uw(\xt)\int d^4y~G_R(x-y)j^{i}_{\rm eff}(y^+,y^- <0, \yt),
\end{equation}
where the "incoming" field $a^i_{\rm in}$ is the solution of the homogeneous fluctuation equation for $x^- <0$ with no sources. Note that the incoming field vanishes since the quark is initially at rest. Now, in order to obtain the radiation field, we have to find the transverse gauge field after the nucleus has passed at $x^- >0$. The corresponding fluctuation equation is easily obtained by noting that $\uw=1$ in \eq\nr{eq:flukeq1}. This yields
\begin{equation}
\label{eq:flukeq2}
\Box a^i  =  j^i_{\rm eff}(x^+,x^- >0, \xt).
\end{equation}
The general solution in terms of the advanced Green function reads 
\begin{equation}
\label{eq:flukeq3}
a^{i}(x^+,x^->0, \xt) = a^{i}_{\rm out} +  \int d^4y~G_A(x-y)j^{i}_{\rm eff}(y^+,y^- >0, \yt),
\end{equation}
where the "out-going" field $a^i_{\rm out}$ is the solution of the homogenous equation $\Box a^i(x^- >0) =0$.

The solution of the transverse radiation field at the surface $x^- = 0$ is then obtained by taking the difference between the out-going and incoming fields
\begin{equation}
\label{eq:radfield}
\begin{split}
a^{i}_{\rm rad}(x^+, x^-=0,\xt) = \uw(\xt)\int d^4y~G_R(x-y)&j^{i}_{\rm eff}(y^+, y^-<0,\yt)\\
& - \int d^4y~G_A(x-y)j^{i}_{\rm eff}(y^+,y^->0,\yt).
\end{split}
\end{equation} 
Note that in taking the difference the left hand sides of \eqs\nr{eq:solflukeq1} and \nr{eq:flukeq3} cancel since both are evaluated at $x^-=0$
and $a_i$ is continuous there. 
Using the notation introduced in \eqs\nr{eq:currentjfull1} and \nr{eq:currentjfull2}, the radiation field at $x^-=0$
with all color indices can be written as 
\begin{equation}
\label{eq:radfieldcolor}
\begin{split}
a^{ia}_{\rm rad}(x^+, x^-=0,\xt) = g\uw_{ad}(\xt)\uw_{db}(\ot)Q^b&\int d^4y~G_R(x-y)\mathcal{J}^{i}_{\rm eff}(y^+, y^-<0,\yt)\\
& -g Q^a\int d^4y~G_A(x-y)\mathcal{J}^{i}_{\rm eff}(y^+,y^->0,\yt),
\end{split}
\end{equation} 
where the source term
\be
\mathcal{J}^{i}_{\rm eff}(y) \equiv \mathcal{J}^{i}(y) - \partial^i \frac{1}{\partial^-}\mathcal{J}^{-}(y)
\label{jiieff}
\ee
and $\mathcal{J}^{\mu}$ for $y^- < 0$ and $y^- >0$ is defined in \eqs\nr{bigj1} and \nr{bigj2}, respectively. 

From \eq\nr{eq:radfieldcolor} we observe that the gluon radiation comes from two different contributions: The first contribution on the right-hand-side in \eq\nr{eq:radfieldcolor} is unique to QCD, and arises from the disturbance of a Coulomb field composed of colored gluons that is disturbed during the collision process $qA \to g$. The second contribution represents the classical electrodynamics (ED)-like gluon radiation caused by the acceleration of the initial
static quark by the nuclear sheet to an unspecied transverse momentum.

In what we are then interested is the radiation field in momentum space. This is given by the Fourier transform 
\be
\label{fouriermom}
a^i_{\rm rad}(k^+,k^-,\kt) = \int d^4x\, e^{-ik\cdot x} a^i_{\rm rad}(x^+, x^-, \xt).
\ee
To construct the full radiation field in momentum space we first note that the radiation field is a free field at $x^- >0$. From this one can conclude that
\be
a_\rmi{rad}^i(k^-,k^ +,\kt)={a_\rmi{rad}^i(k^-,x^-=0,\kt)\over i(k^+-\fra{k_T^2}{2k^-}-i\epsilon)}=
 \frac{-2ik^-}{k^2 + i\epsilon}a^i_{\rm rad}(k^-,\kt),
\label{freefield}
\ee
where $k^-a^i_{\rm rad}(k^-,x^-=0,\kt)$ is the Fourier transform of the boundary value in \eq\nr{eq:radfieldcolor}. Using $k^- \to i\partial_+$ and substituting \eq\nr{freefield} into \eq\nr{fouriermom} yields the result
\be
\label{radfourier2}
ik^2a^{i}_{\rm rad}(k) = \int dx^+ d^2x~ e^{i(k^-x^+ - \kt \cdot \xt)} 2i\partial_{+}a^{i}_{\rm rad}(x^+,x^-=0,\xt).
\ee

Introducing the following notation in \eq\nr{eq:radfieldcolor}
\be
\begin{split}
a^i_{\rm ini}(x^+, x^-=0,\xt) & \equiv \int d^4y~G_R(x-y)\mathcal{J}^{i}_{\rm eff}(y^+, y^-<0,\yt),\\
a^i_{\rm fin}(x^+, x^-=0,\xt) & \equiv  \int d^4y~G_A(x-y)\mathcal{J}^{i}_{\rm eff}(y^+,y^->0,\yt),
\end{split}
\label{greenint}
\ee
considerably simplifies the expression in \eq\nr{radfourier2} yielding the compact expression
\be
\begin{split}
\label{airadfinal}
ik^2a^{ia}_{\rm rad}(k^+,k^-,\kt) = g\uw_{db}(\ot)Q^b  \int d^2x~e^{- i\kt \cdot \xt} \uw_{ad}(\xt) v^i(\xt) -  gQ^a \int d^2x~e^{- i\kt \cdot \xt} u^i(\xt),
\end{split}
\ee
where the fields $v^i$ and $u^i$ are defined as   
\be
\begin{split}
v^i(\xt)& \equiv  2i\int dx^+  e^{ik^-x^+} \partial_{+}a^{i}_{\rm ini}(x^+,x^-=0,\xt), \\
u^i(\xt) & \equiv 2i\int dx^+  e^{ik^-x^+} \partial_{+}a^{i}_{\rm fin}(x^+,x^-=0,\xt).
\end{split}
\ee
Here $v_i$ is sourced by the initial static part of the quark current, likewise $u_i$ is sourced by the recoil quark. Physics of gluon distribution
 is obtained from the square of \eq\nr{airadfinal} and it comes in two parts: The first term describes gluons arising from the interaction of the
 nuclear sheet with the Coulomb field of the initial quark, the second emission from the kicked quark. Interference effects are important and
 will be discussed in the next subsection, see Eq.\nr{intres}.

The Green's function integrals in \eq\nr{greenint} are carried out in detail in Appendix \ref{appA} and lead to the results
\be
\begin{split}
 v^i(\xt)& =  -\frac{i}{2\pi}\frac{x^i}{x_T} 2\sqrt{2}k^-K_1\left (\sqrt{2}k^-x_T\right )= -i\sqrt2 E_i(k^-,\xt),\\
 u^i(\xt)& = -\frac{i}{2\pi}e^{i\frac{k^-}{p^-}\pt \cdot \xt} \frac{x^i}{x_T}\frac{2mk^-}{p^-}K_1\left (\frac{mk^-}{p^-} x_T\right ),
\end{split}\label{vu}
\ee
where $K_1$ is the modified Bessel function of the second kind. Note that the field $v^i \sim \partial_{+}a^{i}_{\rm ini}$ corresponds to color electric field of a static quark in vacuum, where the transverse electric field $E_i$ along $x^-=0$ is generated by the Coulombic vector potential (in $a^- = 0$ gauge) \cite{Kajantie:2019hft}: 
\be
\label{coulombicfield}
F_{+i} = \partial_{+}a_{\rm ini}^i(x^+,0,\xt) \equiv - \frac{1}{\sqrt 2}E^i(x^+,0,\xt) =   -\frac{x^i}{4\pi \sqrt{2} [(x^+)^2/2 + x_T^2 ]^{3/2}}
\ee
with the Fourier transform
\be
E_i(k^-,\xt)=\int dx^+ e^{ik^- x^+}E_i(x^+,0,\xt)={x_i\over x_T}\,{k^-\over \pi}K_1(\sqrt2 k^-x_T).
\label{ECfour}
\ee
For the field $u^i \sim \partial_{+}a^{i}_{\rm fin}$ there would also be a magnetic field.

\subsection{Introducing quantum effects}\label{subquant}

The number spectrum of produced gluons is obtained by considering the relation 
\be
\label{Nproduction}
16\pi^3\frac{dN}{dyd^2k} =  \frac{1}{N} \sum_{\lambda} \bigg\langle \vert \mathcal{M}_{\lambda}(k)\vert^2 \bigg\rangle_{Q,\rho},
\ee
where the gluon four-momentum $k$ is parametrized in terms of momentum space rapidity $y$ and the subscript $\langle \dots \rangle_\rho$ denote an average over the set of color sources.  The classical gluon production amplitude\footnote{Note that we work in the transverse LC gauge where  $a^- =0$.} is given by 
\be
\mathcal{M}_{\lambda}(k) = k^2 a^{i}_{\rm rad}(k)\varepsilon_{\lambda,i},
\ee
where the sum over the gluon transverse polarization vectors is 
\be
\sum_\lambda \varepsilon_{\lambda,i}\varepsilon^{\ast}_{\lambda,j} = \delta_{ij}.
\ee

By substituting \eq\nr{airadfinal} into \eq\nr{Nproduction}, we obtain 
\be
\label{eq:radaverho}
\begin{split}
16\pi^3   {{dN} \over {dy d^2k}} = \frac{g^2}{N} \bigg\langle  \bigg\vert 
\uw_{db}(\ot)Q^b \int d^2x~&e^{- i\kt \cdot \xt} \uw_{ad}(\xt) v^i(\xt) -  Q^a \int d^2x~e^{- i\kt \cdot \xt} u^i(\xt)
 \bigg\vert^2 \bigg\rangle_{Q,\rho} .
\end{split}
\ee
Following the straightforward calculation of the color algebra given in Appendix \ref{appB} yields the result
\be
\label{intres}
\begin{split}
16\pi^3{{dN} \over {dy d^2k}} =  g^2 C_F \int d^2xd^2y~e^{- i\kt \cdot ( \xt-\yt)} \biggl [ & S(\xt -\yt) v^{i}(\xt) v^{i \ast}(\yt)  -    S(\xt ) v^{i}(\xt) u^{i \ast}(\yt)\\
& -    S(\yt) u^{i}(\xt) v^{i \ast}(\yt)  +   u^{i}(\xt) u^{i \ast}(\yt) \biggr ],
\end{split}
\ee
where the color factor $C_F = (N^2-1)/2N$ and the scattering information between the projectile nucleus and the target quark are encoded in the 2-point correlator of two Wilson lines in the adjoint representation
\be
\label{wcorrelators}
\begin{split}
S(\xt -\yt) & \equiv  \frac{1}{N^2-1}  \bigg\langle \tr{(\uw(\xt)\uw^{\dagger}(\yt))}\bigg\rangle_{\rho},\\
S(\xt) & \equiv  \frac{1}{N^2-1}  \bigg\langle \tr{(\uw(\xt)\uw^{\dagger}(\ot))}\bigg\rangle_{\rho},\\
S(\yt) & \equiv  \frac{1}{N^2-1}  \bigg\langle \tr{(\uw(\yt)\uw^{\dagger}(\ot))}\bigg\rangle_{\rho}.
\end{split}
\ee
The explicit calculation of Wilson line correlators in \eq\nr{wcorrelators} within the approximation of the McLerran-Venugopalan (MV) model \cite{mv}, where the 
fluctuations in the nuclear color field sources are Gaussian, is given in Appendix \ref{appC}. This yields the result
\be
S(\xt -\yt) = \exp\biggl [-(Q_s^A)^2D(\xt - \yt)\biggr ],
\ee
where the dipole function $D$ is given by 
\be
D(\xt -\yt) \equiv \int \frac{d^2\pt}{(2\pi)^2}\frac{2}{(\pt^2 + m_{\rm IR}^2)^2}\biggl (1 - e^{i\pt \cdot (\xt -\yt)}\biggr )
\ee
and $m_{\rm IR}$ is an infrared regulator.

Let us then Fourier transform the expression obtained in \eq\nr{intres} to momentum space via convolution in 2-dimensional momentum space defined as 
\be
\mathcal{F}[f \otimes g] \equiv \int \frac{d^2h}{(2\pi)^2} f(\hht - \kt) g(\hht).
\ee
First, consider the term with $S(\xt -\yt)$ in \eq\nr{intres}. By introducing the following change of variables, $\xt = \bt + \frac{1}{2}\rt$ and $\yt = \bt - \frac{1}{2}\rt$, we obtain 
\be
\int d^2r\, S(\rt)\,e^{-i\kt\cdot\rt}\int d^2b\,v^i(\bt+\fra12\rt)v^{i \ast}(\bt-\fra12 \rt).
\ee
Since $S$ only depends on $\rt$ the $\bt$-integration can be brought to the $v^i v^{i \ast}$ terms and carried out with two different Fourier momenta. This yields the result 
\be
\int {d^2q\over(2\pi)^2}S(\kt - \hht)v^i(\hht)v^{i \ast}(\hht),
\ee
where the Fourier transforms of $v^i$ and $u^i$ are given in Appendix \ref{appA}.  The second and the third term in \eq\nr{intres} follows similarly by considering the Fourier transforms of $S,v$ and $u$. This yields the result
\be
\,\int d^2x e^{-i\kt\cdot\xt}S(\xt)v^i(\xt)\,\int d^2y\, e^{i\kt\cdot\yt}u^{i \ast}(\yt)= \int {d^2h\over(2\pi)^2}S(\kt-\hht)v^i(\hht)u^{i \ast}(\kt),
\ee
and 
\be
\,\int d^2x e^{-i\kt\cdot\xt}u^i(\xt)\,\int d^2y\, e^{i\kt\cdot\yt}S(\yt)v^{i*}(\yt)= \int {d^2h\over(2\pi)^2}S(\kt-\hht)v^{i \ast}(\hht)u^i(\kt).
\ee
Finally, the last term is directly a product $u^i(\kt)u^{i \ast}(\kt)$ of Fourier transforms, but using the normalisation condition 
\be
 \int \frac{d^2h}{(2\pi)^2} S(\hht) = 1
 \label{Snorm}
\ee
it can be combined with the three other terms and one can write the outcome in the form 
\be
16\pi^3{dN\over dyd^2k}\nn = g^2C_F \int{d^2h\over (2\pi)^2}S(\kt-\hht)\, \biggl (v^i(\hht)-u^i(\kt)\biggr )\biggl (v^{*i}(\hht)-u^{*i}(\kt)\biggr ).
\label{square}
\ee

By substituting the explicit expression for $v$ and $u$ (see Appendix \ref{appA} \eqs\nr{uki} and \nr{vki}), we find a very compact and elegant expression for the gluon radiation spectrum 
\be
\label{finresspectra}
\begin{split}
\left ({16\pi^3\over g^2C_F}\right ) {{dN} \over {dy d^2k}} = 2\int \frac{d^2h}{(2\pi)^2} S(\kt -\hht) \biggl [\frac{h^i}{h^2_T + 2(k^-)^2} - \frac{k^i - \xi p^i}{\vert \kt - \xi \pt\vert^2 + \xi^2 m^2   }\biggr ]^2,\\
\end{split}
\ee
where we have defined $\xi \equiv k^-/p^-$. This is the main result of our work.

\subsection{Discussion of the main result in \eq\nr{finresspectra}}
\label{resdisc}

\subsubsection{Relation to known results}\label{subsubrelation}

As a first check one can compare with known results in the central region, which corresponds to $y \to \infty$ in \eq\nr{finresspectra}. In this limit, $k^-\sim e^{-y}\to0$, $\xi=k^-/p^-\to0$,
and the bracketed factor becomes
\be
\left({h^i\over h_T^2}-{k^i\over k_T^2}\right)^2 = {|\kt-\hht|^2\over h_T^2 k_T^2}.
\ee
This shows immediately the relation to Eq. (23) of \reff\cite{gelism-t}, where the gluon production amplitude is given for $p+A$ collisions 
in a similar temporal LC gauge. The RHS of equation \nr{finresspectra} thus becomes 
\be
\label{bulkspectracent}
\begin{split}
\left ({16\pi^3\over g^2C_F}\right ){{dN} \over {dy d^2k}}\bigg\vert_{{\rm bulk},y\to\infty} = 
2\int \frac{d^2h}{(2\pi)^2} S(\kt -\hht) \frac{\vert \kt - \hht\vert^2}{h_T^2k_T^2},
\end{split}
\ee
reminiscent of known results for gluon production in $p+A$ in \reff\cite{blaizotgv} (the color correlator on the proton side is missing). 
More specifically, using for $S$ the standard large momentum approximation $k_T \gg Q^{A}_s$  (see the derivation in \eq\nr{gp} in Appendix  B) 
\be
\label{largeS}
S(\kt) = \frac{2Q_s^2}{k_T^4} + \mathcal{O}\left (\frac{Q_s^4}{k_T^6}\right ),
\ee
where the non-linearities of the projectile color field are unimportant, the RHS in \eq\nr{bulkspectracent} becomes 
a Gunion-Bertsch-type \cite{gb} infrared divergent spectrum in the central region
\be
\label{bulkspectracentGB}
\begin{split}
\left ({16\pi^3\over g^2C_F}\right ){{dN} \over {dy d^2k}}\bigg\vert_{\rm bulk} \approx  \frac{4(Q^A_s)^2}{k_T^2}\int \frac{d^2h}{(2\pi)^2}  \frac{1}{\vert \kt - \hht\vert^2h_T^2}. 
\end{split}
\ee

\subsubsection{Separation of bulk and bremsstrahlung}\label{subsubseparation}
To interpret the structure of the result turns out to be useful to add and subtract 
\be
{1\over\sqrt2}iE(k^-,\kt)={k^i \over k^2_T + 2(k^-)^2}
\ee 
inside the the brackets in \eq\nr{finresspectra}. This yields the result
\be
\label{finresspectrav2}
\begin{split}
\left ({16\pi^3\over g^2C_F}\right ){{dN} \over {dy d^2k}} = 2\int \frac{d^2h}{(2\pi)^2} S(\kt -\hht) \biggl [\mathcal{M}_{\rm bulk}^i + 
\mathcal{M}_{\rm brems}^i\biggr ]^2,
\end{split}
\ee
where we have introduced the compact notation 
\be
\mathcal{M}_{\rm bulk}^i \equiv \frac{h^i}{h^2_T + 2(k^-)^2}  - \frac{k^i}{k^2_T + 2(k^-)^2},
\label{bu}
\ee
and 
\be
\mathcal{M}_{\rm brems}^i \equiv \frac{k^i}{k^2_T + 2(k^-)^2}  - \frac{k^i - \xi p^i}{\vert \kt - \xi \pt \vert^2 + \xi^2 m^2}.
\label{bre}
\ee
What we have hereby gained is a separation of scales. The bulk term, denoted as $\mathcal{M}^i_{\rm bulk}$, now depends on the convolution and the produced gluon momenta, there is no dependence on the kicked quark
momentum. The bremsstrahlung term, denoted as $\mathcal{M}_{\rm brems}^i$, on the other hand, is completely independent of the convolution momentum so that, due to the normalisation 
of $S(q)$ in \eq\nr{Snorm}, the integration over $h^i$ can be carried out and $S(q)$ and all nuclear effects are entirely eliminated. Since $\xi=k^-/p^-$ and
$k^-\sim e^{-y}$ it also manifestly vanishes in the central region where $y$ is large.

In the following section, we will address the three separate contributions to the gluon radiation spectrum obtained in \eq\nr{finresspectrav2}. 
The first comes from $|\mathcal{M}_{\rm bulk}^i|^2 $,  and it is the contribution from the the interaction between the nuclear sheet and the electric field of the static initial quark.
The second is the bremsstrahlung contribution coming from $|\mathcal{M}_{\rm brems}^i|^2$. Taken literally as it stands in \eq\nr{bre} it contains 
no dependence on the nucleus. It gives precisely classical ED gluon emission distribution going $\sim1/k_T^2$ even at asymptotic $k_T$. 
We shall show that this is due to the assumed constant kick and has to be corrected by a momentum $q$ absorbed from the nucleus. Finally, the third contribution comes from the possible interference contribution between the bulk and bremsstrahlung.

%% file: corrbbiv2.tex
\section{Properties and numerics of bulk and bremsstrahlung}
\subsection{Bulk contribution}\label{subbulk}
Let us first consider the bulk contribution in \eq\nr{finresspectrav2} which, after opening the brackets, yields the expression
\be
\label{bulkspectra}
\begin{split}
\left ({16\pi^3\over g^2C_F}\right ){{dN} \over {dy d^2k}}\bigg\vert_{\rm bulk} & \equiv 2\int \frac{d^2h}{(2\pi)^2} S(\kt -\hht) \biggl [ \frac{h_T^2}{(h_T^2 + 2(k^-)^2)^2} \\
&  - \frac{2\hht \cdot \kt}{(h_T^2 + 2(k^-)^2)(k_T^2 + 2(k^-)^2)}  + \frac{k_T^2}{(k_T^2 + 2(k^-)^2)^2}  \biggr ],
\end{split}
\ee
where $2(k^-)^2 = k_T^2e^{-2y}$. Very schematically, one could describe the terms in \eq\nr{bulkspectra} as follows. The first term
is the square of the amplitude of a process in which the target quark emits a gluon of momentum $\hht$ which then goes through
the nuclear sheet, picks up a momentum $\kt-\hht$ and produces a gluon of momentum $\kt$. All momenta $\hht$ are
integrated over.  This amplitude is the first term in \eq\nr{eq:radaverho} and it 
is what one would automatically write down, but this would lead to physically wrong conclusions. 
For example, when one computes the leading term at large $k_T$, it could be negative for some rapidities. There is
another process, in which the quark emits before the collision. This is the second term in \eq\nr{eq:radaverho} but with $u^i(\xt)$ replaced
by $v^i(\xt)$, the initial state emission function (``emitting a gluon of momentum $\kt$'' means taking the Fourier transform over $d^2x$ 
of this function). This process interferes with the first one and produces the second term in
\eq\nr{bulkspectra}, its square gives the last term. Including all of these eliminates all non-physicalities.

Let us then study the large $k_T$ behaviour of \eq\nr{bulkspectra} in the case of finite gluon rapidity $y$, which is $\mathcal{O}(1)$ in the target fragmentation region we are interested in. Applying again the expansion in \eq\nr{largeS} leads to 
\be
\label{bulkspectralargeS}
\begin{split}
 &\left (\frac{16\pi^3}{g^2C_F} \right ) {{dN} \over {dy d^2k}}\bigg\vert_{\rm bulk} =   +  \frac{(Q^A_s)^2}{\pi m_{\rm IR}^2 k_T^2(1+ e^{-2y})^2}   +\int \frac{d^2h}{(2\pi)^2} \frac{4(Q^A_s)^2}{(\vert \kt -\hht\vert^2 + m^2_{\rm IR})^2} \\
 & \times \biggl [ \frac{h_T^2}{(h_T^2 + k_T^2e^{-2y}+ m^2_{\rm IR})^2} -
  \frac{2\hht \cdot \kt}{(h_T^2 + k_T^2e^{-2y}+ m^2_{\rm IR})(k_T^2 +k_T^2 e^{-2y})}  \biggr ] + \mathcal{O}\left (\frac{(Q^A_s)^2}{k_T^2}\right ),
\end{split}
\ee
where the first term in RHS of \eq\nr{bulkspectralargeS} arises from the trivial $h$-integration over the constant $ k_T^2/(k_T^2 + k_T^2e^{-2y})^2$ contribution and the mass term $m_{\rm IR}$ is introduced as an infrared regulator. Of course, for small $y$, 
as long as $k_T^2e^{-2y}>m_\rmi{IR}^2$, it is not needed in two last terms, but becomes important for
\be
y\ge \log{k_T\over m_\rmi{IR}},
\label{brlimit}
\ee
and, in fact, has important dynamical consequences for the $y$ distribution. 
 
In order to get the full gluon spectrum at large $k_T$ we first carry out exactly the integrals over azimuthal angle and then over $h_T$ and finally expand in small $m_{\rm IR}$ (in the $(\vert \kt -\hht\vert^2 + m^2_{\rm IR})$ term). This gives the result
\be
\label{bulkspectralargeSfull}
\begin{split}
 \left (\frac{16\pi^3}{g^2C_F} \right ) {{dN} \over {dy d^2k}}\bigg\vert_{\rm bulk} =  
 \frac{2}{\pi (1+e^{-2y})^2}&\frac{(Q^A_s)^2}{k_T^4}\biggl [\frac{1}{2}\left (1 + \tanh^2(y) \right )
 \log \left (\frac{k_T^2(1+e^{-2y})}{m_{\rm IR}\sqrt{m^2_{\rm IR}+k_T^2e^{-2y}}} \right )\\
 & - \tanh^2(y) \biggr ] + \mathcal{O}\left (m^2_{\rm IR},\frac{(Q^A_s)^2}{k_T^2}\right ),
\end{split}
\ee
where the powerlike divergence $\sim 1/(k_Tm_{\rm IR})^2$ in \eq\nr{bulkspectralargeS} cancels out after the integration. As expected, we recover the standard perturbative $\sim 1/k_T^4$ behavior with a logarithmic correction. The overall rapidity structure can be written in the more explicit form
\be
\begin{split}
&\left (\frac{16\pi^3}{g^2C_F} \right ) {{dN} \over {dy d^2k}}\bigg\vert_{\rm bulk} =
  \frac{2(Q^A_s)^2}{\pi k_T^4}  \frac{1}{ (1+e^{-2y})^2}\\
& \hspace{-3mm}\times \biggl [\frac{1}{2}\left (1 + \tanh^2(y) \right )\left(y+\log(1+e^{-2y})-\fra12\log(1+{ m_\rmi{IR}^2\over k_T^2}e^{2y})+\log{k_T\over m_\rmi{IR}}\right) - \tanh^2(y) \biggr ].
\label{ydepeq}
\end{split}
 \ee
This expression emphasises the fact that at small  $y<\log(k_T/m_\rmi{IR})$ the distribution has a dominant linear increase in $y$, which above this turns
to a plateau.

\begin{figure}[!t]
\begin{center}
\includegraphics[width=0.6\textwidth]{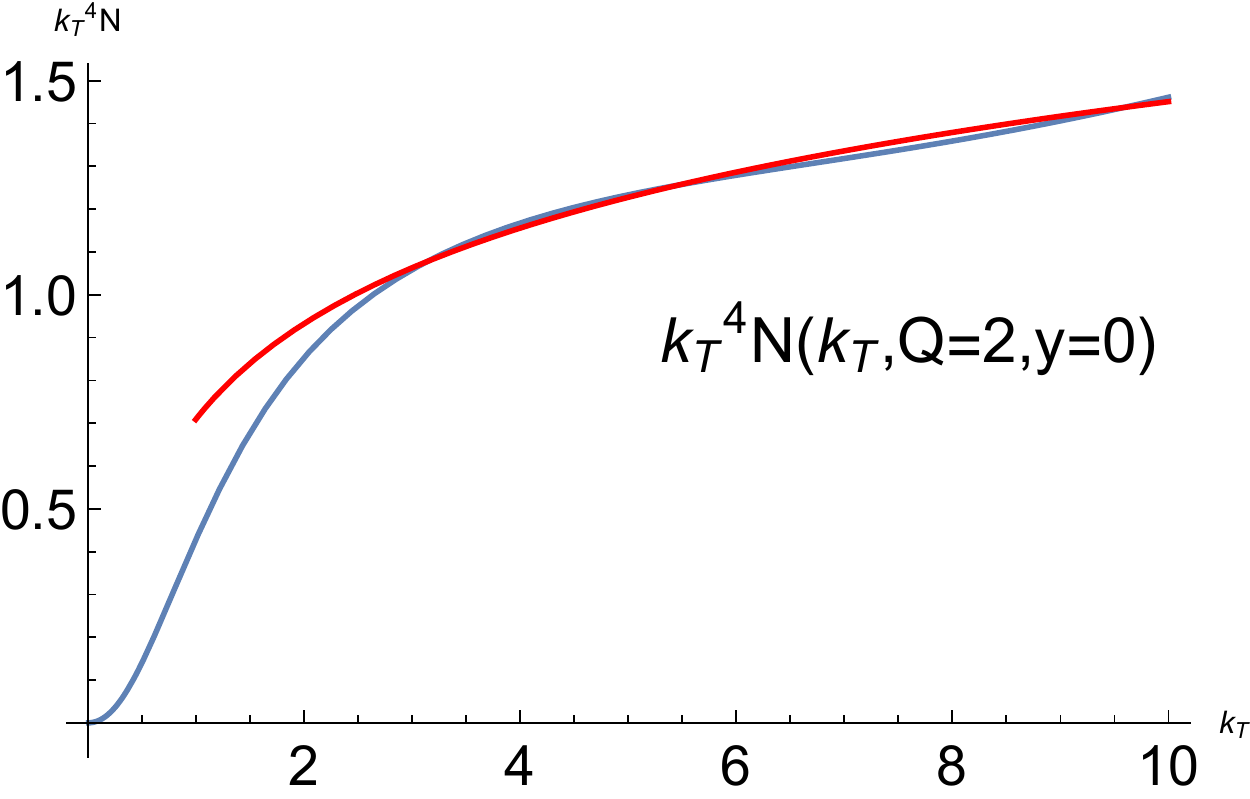}
\end{center}
\vspace{-0.8cm}
\caption{\small A plot of $k_T^4 N(k_T,y=0)$ computed from \eq\nr{bulkspectranum} for 
$y=0$ and for the correlator $S$ computed for $Q^A_s=2$ and $m_{\rm IR} =0.2$. The red curve is the logarithmic
approximation in \eq\nr{loga}. 
}
\label{NkTfig}
\end{figure}

\begin{figure}[!t]
\begin{center}
\includegraphics[width=0.48\textwidth]{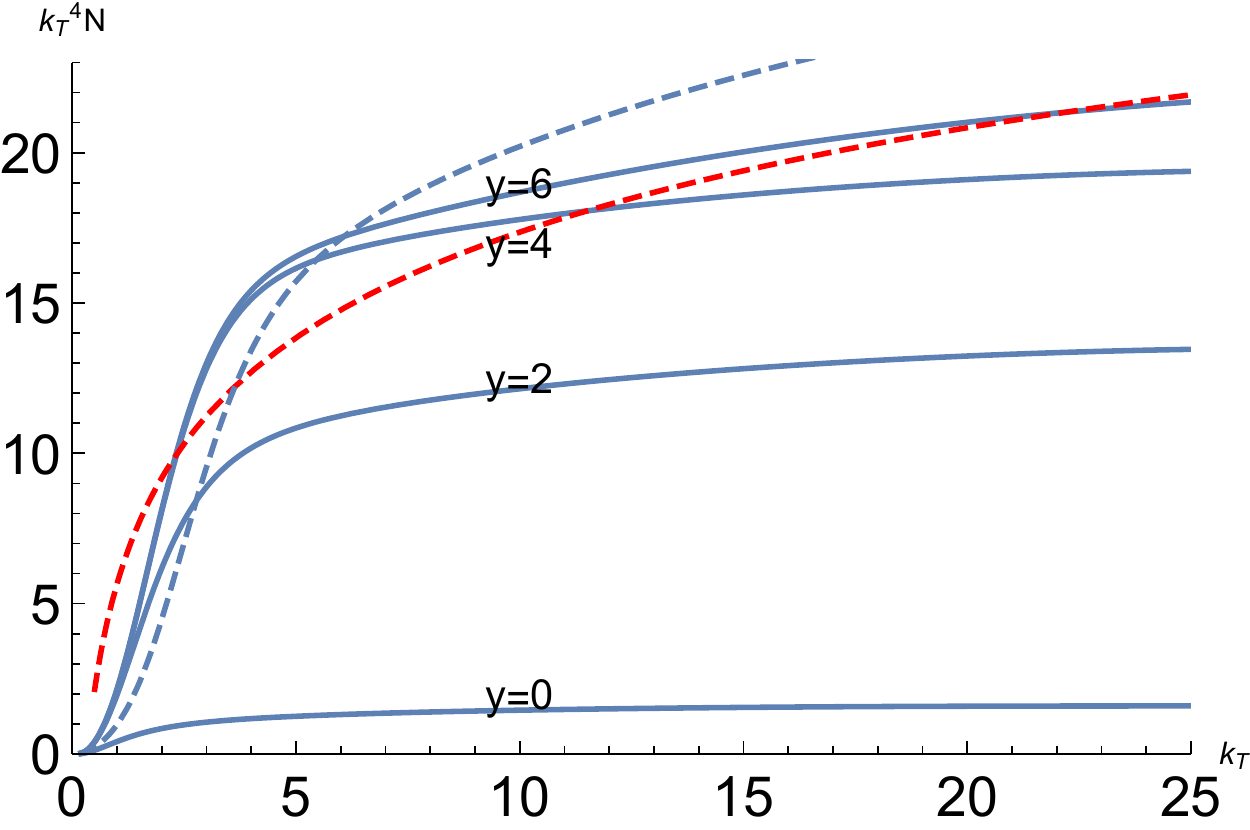}
\includegraphics[width=0.48\textwidth]{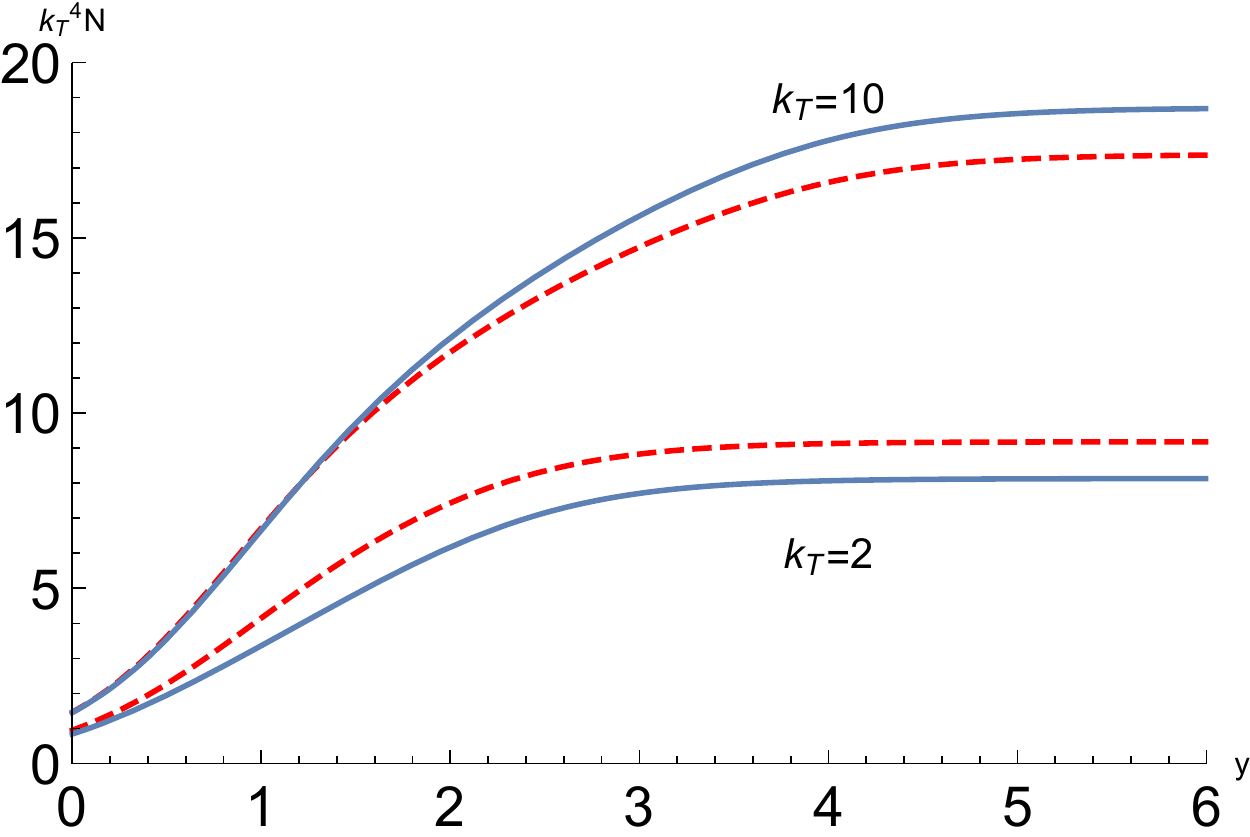}
\end{center}
\caption{\small A plot of $k_T^4 N(k_T,y)$ computed from \eq\nr{bulkspectranum} for values of 
$y$ maked in the figure and for $Q^A_s=2$ and $m_{\rm IR} =0.2$ (left panel). The red dashed curve is the analytic large $k_T$
approximation \eq\nr{ydepeq} for $y=6$. The blue dashed curve is the $y\to\infty$ limit \eq\nr{dNCR}.
The right panel shows the $y$ dependence for $k_T=2, 10$ and for the same $Q_s^A,\,m_{\rm IR}$. The red dashed curve is the analytic approximation in \eq\nr{ydepeq}. The curves are proportional to  $(Q^A_s)^2$.
}
\label{yfig}
\end{figure}

Consider then full numerical evaluation of the bulk result in \eq\nr{bulkspectra}. To evaluate the $h$-integral we first make the change of variables, $\hht = \hht - \kt$, and do analytically  the azimuthal integral around $\kt$. The $\hht$ integral in \eq\nr{bulkspectra} becomes 
\be
\label{bulkspectranum}
\begin{split}
\left ({16\pi^3\over g^2 C_F}\right ){{dN} \over {dy d^2k}}\bigg\vert_{\rm bulk} = 2 \int_{0}^{\infty} d\hat h_T \frac{\hat h_T S(k_T \hat h_T)}{2\pi} w_{\rm bulk}[\hat h_T,y],
\end{split}
\ee
where the variable $k_T$ has been scaled out by using $h_T \equiv k_T\hat h_T$ and the weight function $w$ is given by 
\be
\label{wfunc}
\begin{split}
w_{\rm bulk}[\hat h_T,y] = & \frac{(1+e^{-2y} + \hat h_T^2)(1 + \hat h_T^2) - 4\hat h_T^2}{[(1+e^{-2y} + \hat h_T^2)^2 - 4\hat h_T^2]^{3/2}}  + \frac{1}{(1 + e^{-2y})^2}\\
& - \frac{1}{1 + e^{-2y}}\biggl [\frac{1 - e^{-2y} - \hat h_T^2}{[(1+e^{-2y} + \hat h_T^2)^2 - 4\hat h_T^2]^{1/2}}  + 1 \biggr ]. 
\end{split}
\ee
The expression in \eq\nr{bulkspectranum} can  be integrated numerically by using \eq\nr{Sfourier} to numerically compute the Fourier transform $S(q_T)$ of the correlation function $S(r_T)$. 
For $y=0$ this yields the $k_T$ spectrum shown in \fig\ref{NkTfig}. Curves always refer to the integral in the right hand side of \eq\nr{bulkspectranum}, without the factors $g^2C_F$. From \fig\ref{NkTfig}, one observes scaling with $k_T^4$, but also a suggestion of logarithmic variation. This can be fitted by
\be
0.322 \log(k_T/2) +0.932,
\label{loga}
\ee
which is plotted in red in \fig\ref{NkTfig}. This quantitatively agrees with the logarithmic factor
\be
\frac{1}{\pi}\log(2k_T/m_{\rm IR}) \sim 0.318 \log(k_T/2) + 0.953
\ee
from \eq\nr{bulkspectralargeSfull} with $Q^A_s=2, y=0$ and $m_{\rm IR} =0.2$. 

In \fig\ref{yfig} we show (left  panel)  the $k_T^4$-scaled $k_T$ spectrum of gluons as a function of 
$k_T$ for various fixed rapidities $y=0,1,2,4,6$ or (right panel)
as a function of $y$ plotted for $k_T=2,\,10$, together with the large $k_T$ approximation \nr{ydepeq} 
(red dashed curves in both panels).

The rapidity distribution in the fragmentation region shows first at small $y$ some structure 
which is well reproduced by the approximation  \eq\nr{ydepeq}, then a quasilinear increase which at larger $y$
turns to a plateau. The key factor producing this pattern is  the scale of the IR log in \eq\nr{bulkspectralargeSfull}.
At small $y$ the 
IR divergence of the $1/(h_T^2+2(k^-)^2)$ term is cut off by $2(k^-)^2=k_T^2\,e^{-2y}$. This disappears at large $y$
and the boundary clearly is when it decreases below $m_\rmi{IR}^2$, i.e. IR screening by $m_\rmi{IR}^2$ in all terms is effective (compare \eq\nr{bulkspectralargeSfull}) as long as
\be
y\le \log{k_T\over m_\rmi{IR}}.
\label{brlimit2}
\ee
We will later identify this as the lower limit of the bremsstrahlung contribution. We do not see evidence of an upper limit of the
fragmentation region, there is just the merging into the central plateau. In contrast, we shall find that the bremsstrahlung contribution decays
as $\sim e^{-2y}$ when one approaches to central region (see \eq\nr{bremssfull}).

\begin{figure}[!t]
\begin{center}
\includegraphics[width=0.48\textwidth]{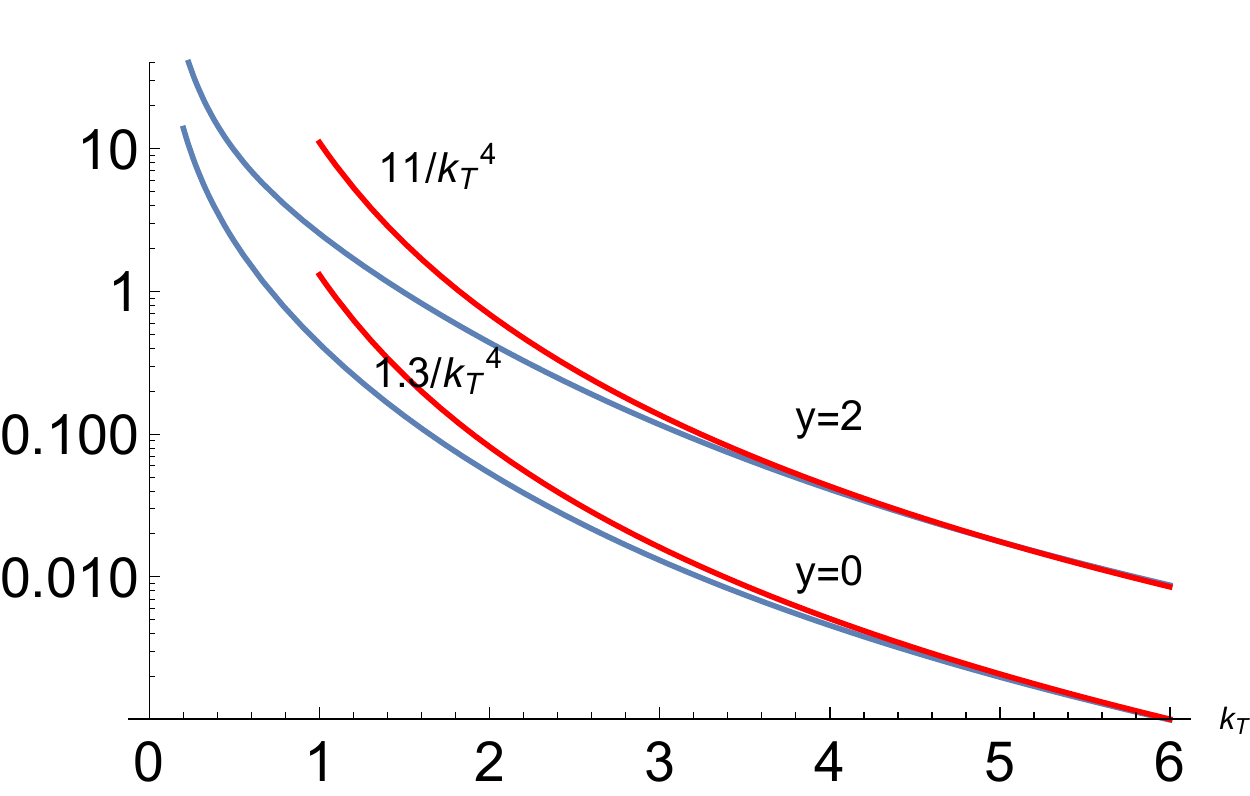}
\includegraphics[width=0.48\textwidth]{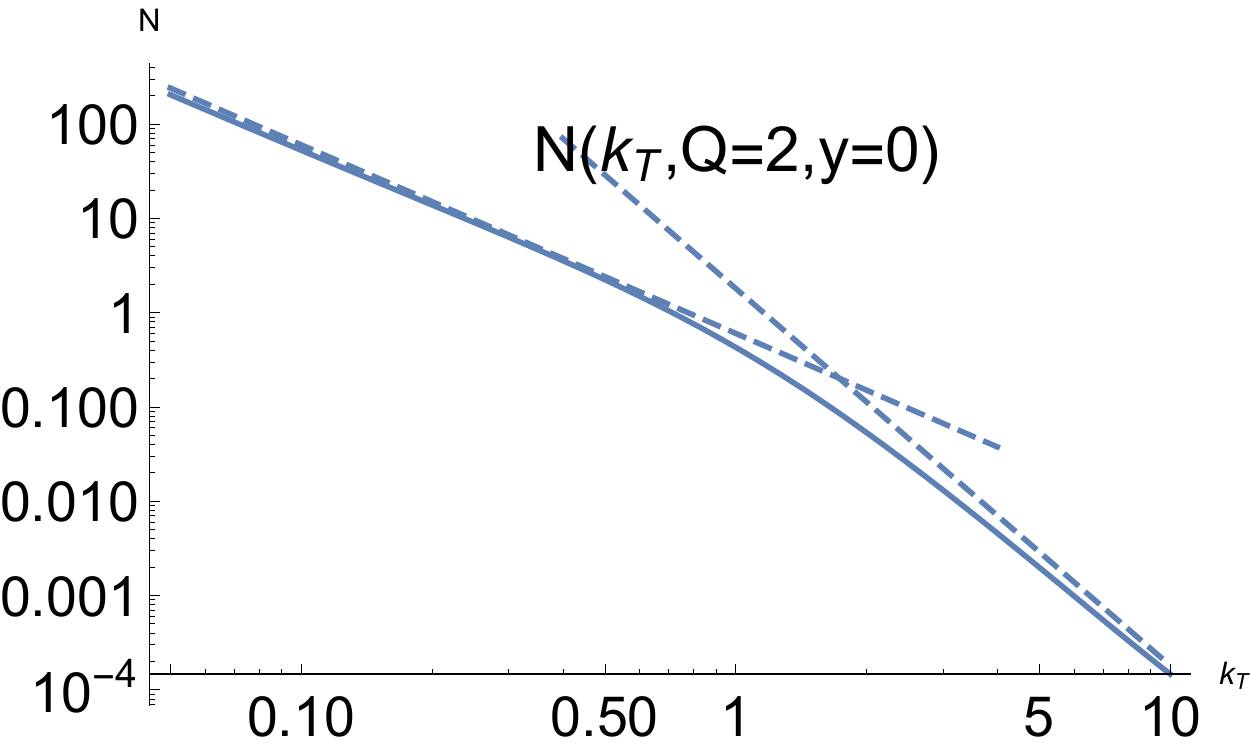}
\end{center}
\caption{\small A plot of unscaled $N(k_T,y)$ for small $k_T$ and $y=0,2$. The red curves are powerlike
approximations ${\rm const}/k_T^4$.  The right panel shows the data for $y=0$ on a log-log scale so that a transition from
$1.3/k_T^4$ to $0.6/k_T^2$ (dashed curves) is visible.
}
\label{NkTfig2}
\end{figure}

The limiting curve at large $y$ is simply obtained by taking this limit in the integrand or of the weight function \eq\nr{wfunc}. The result is
\be
{2\over k_T^2}\int_0^\infty dh_T{h_TS(h_T)\over 2\pi}\,{h_T^2\over |k_T^2-h_T^2|} = 
{2\over k_T^2}\int {d^2h\,S(h_T)\over (2\pi)^2}\,{h_T^2\over |k_T^2-h_T^2|}.
\label{dNCR}
\ee
Regularising the pole at $h_T=k_T$ in the range of integration by replacing the denominator above by
$\sqrt{(k_T-h_T)^2+(Q^A_s)^4}$, $Q^A_s=2$ one obtains the central region $k_T$ distribution, multiplied by $k_T^4$,  plotted in Fig.\ref{yfig}. 
One sees that in the target fragmentation region there is a slight enhancement relative to the $k_T$ distribution in the central
region. The details of the computation, especially the extrapolation of the central region distribution this far in $y$, are very
uncertain.
Phenomenologically, by multiplying by $k_T$, this would give an estimate of how much a single quark contributes to energy deposition in
the target fragmentation region.

In \fig\ref{NkTfig2}, we show the numerically computed  gluon distribution without $k_T$-scaling so that one sees what happens in the saturation
region, $k_T \ll Q^A_s =2$. The perturbative $1/k_T^4$ behavior at large $k_T$ slows to a $1/k_T^2$ behavior, but this persists
down to $k_T\to 0$ since there is no target saturation scale. In fact, this is due to the second term in \eq\nr{bulkspectra}, which is precisely the ED-like radiation from quark acceleration.
In a study of $p+A$ collisions there, qualitatively, would be a second saturation scale $Q^p<Q_s^A$ which would shield the $1/k_T^2$
behavior at small $k_T$.

\subsection{Bremsstrahlung and its interference with bulk}
\label{bremint}

We have above split the general result \eq\nr{finresspectrav2} in the bulk part \eq\nr{bu}, which contains all nuclear effects and is independent
of the assumed constant kick of the quark and the bremsstrahlung part \eq\nr{bre}, which does not contain nuclear effects, depends sensitively
on the magnitude of the kick and is bounded from below (as we shall show) and above in rapidity.
In fact, it is the same as classical ED radiation due to sudden acceleration of charged particle.

Inserting in \eq\nr{bre} the variable $v_T=p_T/\sqrt{p_T^2+m^2}$ with 
\be
\begin{split}
\label{kinematics1}
k_T^2+2(k^-)^2 & = k_T^2(1+e^{-2y}),\\
\vert \kt - \xi \pt \vert^2 & =k_T^2(1-2v_T e^{-y+y_p}\cos\phi+v_T^2 e^{-2y+2y_p}),
\end{split}
\ee
and
\be
\begin{split}
\label{kinematics2}
 \vert \kt - \xi \pt \vert^2+\xi^2m^2 & = k_T^2(1-2v_T e^{-y+y_p}\cos\phi+e^{-2y+2y_p}), \\
 \kt\cdot(\kt-\xi\pt)& =k_T^2(1-v_T \,e^{-y+y_p}\cos\phi ),
\end{split}
 \ee
 gives the result
 \be
\label{bremssfullv2}
\begin{split}
 \left (\frac{16\pi^3}{g^2C_F} \right ) {{dN} \over {dy d^2k}}\bigg\vert_{{\rm bremss}} = & \frac{2}{k_T^2}\biggl [\frac{1}{(1+e^{-2y})^2}  + \frac{1-2v_Te^{-(y-y_p)}\cos\phi + v_T^2e^{-2(y-y_p)}}{(1-2v_Te^{-(y-y_p)}\cos\phi + e^{-2(y-y_p)})^2}\\
 & - \frac{2(1-v_Te^{-(y-y_p)}\cos\phi)}{(1+e^{-2y})(1-2v_Te^{-(y-y_p)}\cos\phi + e^{-2(y-y_p)})} \biggr ],
 \end{split}
 \ee
 where we have defined $\cos \phi \equiv \kt \cdot \pt/(k_Tp_T)$. From \eq\nr{bremssfullv2} one sees that a factor $1/k_T^2$ scales out of $\vert \mathcal{M}_{\rm brems}\vert^2$, and with fixed kick the entire bremsstrahlung gluon distribution is $1/k_T^2$ times a function of $y,\,y_p$, $p_T$, $\phi$ and $m$. For small $m$ the entire contribution 
of \eq\nr{bremssfullv2} to the gluon spectrum simplifies to (note the extra factor 2)
\be
\label{bremssfull}
 \left (\frac{16\pi^3}{g^2C_F} \right ) {{dN} \over {dy d^2k}}\bigg\vert_{{\rm bremss},\,m=0}=
 {1\over 2k_T^2}{\cosh(y+y_p)+\cos\phi\over  \cosh^2y\,(\cosh(y-y_p)-\cos\phi)}.
 \ee
 At large $y$ this decreases $\sim e^{-2y}$,  which implies that bremsstrahlung contribution vanishes in the central region.
In the denominator one has $p_\rmi{in}\cdot k=-mk_T\cosh y$, $p\cdot k=-k_Tp_T(\cosh(y-y_p)-\cos\phi)$ so that there is
a collinear singularity for emission parallel to the kicked quark. This is shielded by finite quark mass $m$. 
We are mainly interested in $y$ dependence for $y$ close to zero and for $y_p$ large, large recoil. Its magnitude then is order $1/k_T^2$, there is no
dependence on the nucleus so no nuclear enhancement by a factor $(Q^A_s)^2$ like for the bulk contribution. Even more importantly,
the general dependence $\sim 1/k_T^2$ cannot hold to very large $k_T$: ultimately a gluon propagator squared $\sim 1/k_T^4$
must enter. The source of this deficiency at large $k_T$ is the assumption of constant kick or constant momentum transfer
from the nucleus.

There are kinematic constraints on the process 
\be
P_A + p_\rmi{in} = P_X+p+k,\quad P_A=P_X+q,\quad q+p_\rmi{in}=p+k,
\label{kin}
\ee
studied. Since in the thin sheet limit $P_A=(P^+_A\to\infty,0,{\bf 0})$ we have also $P_X^+\to\infty$ from conservation of the + component.
Then the minus components $P_A^-=M_A^2/(2P_A^+)$ and $P_X^-=M_X^2/(2P_X^+)\approx -q^-$ are also very small. For usual mass shell
particles $q^-$ is $\le0$. The conservation rules of minus and transverse components become
\be
q^-+\fra{m}{\sqrt2}=p^-+k^-,\quad \qt=\pt+\kt = -{\mathbf{P_{\rm X}}}.
\label{kinlim}
\ee
For the minus components  
\be
\sqrt2 q^-+m = k_T\,e^{-y}+m_T e^{-yp}.
\ee
The RHS is positive so sufficiently large negative $q^-$ can even make the process impossible for on-shell particles. Assuming $q^-=0$
one can solve
\be
y=\log{k_T\over m-m_T e^{-y_p}} >\log{k_T\over m}.
\ee
We are thus back to \eq\nr{brlimit2} and see that the lower limit of bremsstrahlung is approximately the same 
as the estimated lower limit of the rapidity plateau of the bulk distribution in the MV model.  To reach large $k_T$ 
one must go to the central region, $y\gg1$. Conversely, at fixed rapidity, there is an upper limit to $k_T$, $k_T<m e^y$. This will cut
off the $1/k_T^2$ distribution. 

The bulk contribution, i.e., the MV model, produces in the fragmentation region the increasing distribution in \fig\ref{yfig},
proportional to $(Q^A_s)^2$ at large $k_T$. Due to many uncertainties, the detailed comparison of relative magnitudes  at large $k_T$ will have to wait for a definite model of improved bremsstrahlung. Anyway bremsstrahlung will decay $\sim e^{-2y}$ with increasing $y$ and in the central region the gluonic bulk will dominate.

\subsection{Interference contribution}
\label{interference}

Let us then consider interference term between the bulk and bremsstrahlung contributions
\be
\label{interf}
\begin{split}
\left ({16\pi^3\over g^2C_F}\right ){{dN} \over {dy d^2k}}\bigg\vert_{\rm interf}  \equiv 4\int \frac{d^2h}{(2\pi)^2} & S(\kt -\hht) \biggl [ \frac{h^i}{h_T^2 + 2(k^-)^2} - \frac{k^i}{k_T^2 + 2(k^-)^2}\biggr ] \\
&  \times \biggl [\frac{k^i}{k_T^2 + 2(k^-)^2}  - \frac{k^i - \xi p^i}{\vert \kt - \xi \pt\vert^2 + \xi^2 m^2} \biggr ].
\end{split}
\ee
To do the transverse momentum $h_T$ integral numerically and to do one angular integral analytically, we have to switch the
convolution factor $\kt - \hht$ to the brackets so that the integral to be done becomes
\be
\label{interf2}
\begin{split}
 \left ({16\pi^3\over g^2C_F}\right )&{{dN} \over {dy d^2k}}\bigg\vert_{\rm interf}  =  4\int \frac{d^2h}{(2\pi)^2}  S(\hht) \biggl [ \frac{k_T^2 - \hht \cdot \kt}{(\vert \kt - \hht\vert^2 + a^2)(k_T^2 + a^2)}\\
&  - \frac{k_T^2 - \hht \cdot \kt - \xi (\kt \cdot \pt  - \hht \cdot \pt)}{(\vert \kt - \hht\vert^2 + a^2)(\kappa^2 + \xi^2 m^2)} - \frac{k_T^2}{(k_T^2 + a^2)^2} + \frac{k_T^2 - \xi \kt \cdot \pt}{(k_T^2 + a^2)(\kappa^2 + \xi^2 m^2)}\biggr ],
\end{split}
\ee
where we abbreviated $\kappa^2 \equiv \vert \kt - \xi \pt\vert^2$, $a^2 \equiv 2(k^-)^2$ and the variable $\xi = (k_T/m_T)e^{-y + y_p}$.  The integration over the angle
between $\kt$ and $\hht$ can now be carried out and the integral yields
\be
\label{interf3}
\begin{split}
 & \left ({16\pi^3\over g^2C_F}\right ){{dN} \over {dy d^2k}}\bigg\vert_{\rm interf}  =  2\int_{0}^{\infty} dh_T \frac{h_T S(h_T)}{2\pi} \biggl [\frac{k_T^2 - a^2 - h_T^2 + \sqrt{K}}{\sqrt{K}} \biggl \{\frac{1}{k_T^2 + a^2}\\
 &  - \frac{1}{\kappa^2 + \xi^2 m^2}\left (1 - \frac{p_T}{k_T}\xi \cos(\phi) \right ) \biggr \} - \frac{2k_T^2}{(k_T^2 + a^2)^2} + \frac{2k_T^2 - 2\xi k_T p_T \cos(\phi)}{(k_T^2 + a^2)(\kappa^2 + \xi^2 m^2)}\biggr ],
\end{split}
\ee
where we have defined the factor $K$ as 
\be
\label{factorK}
K \equiv (a^2 + k_T^2 + h_T^2)^2 - 4k_T^2h_T^2.
 \ee
Using the kinematics introduced in \eqs\nr{kinematics1} and \nr{kinematics2} and scaling out the variable $k_T$ by denoting $h_T = k_T \hat h_T$ the expression in \eq\nr{interf3} simplifies to 
\be
\label{interf4}
\begin{split}
 & \left ({16\pi^3\over g^2C_F}\right ){{dN} \over {dy d^2k}}\bigg\vert_{\rm interf}  =  2\int_{0}^{\infty} d\hat h_T \frac{\hat h_T S(k_T \hat h_T)}{2\pi} \omega_{\rm interf}[\hat h_T; v_T, y, y_p, \phi],
\end{split}
\ee
where the weight function $\omega_{\rm interf}$ is given by 
\be 
\begin{split}
\omega_{\rm interf}[\hat h_T; v_T, y, y_p, \phi] & = \biggl [\frac{1 - e^{-2y}- \hat h_T^2}{[(1+e^{-2y} +\hat h_T^2)^2 - 4\hat h_T^2 ]^{1/2}} - \tanh(y) \biggr ]\\
& \times \biggl \{\frac{1}{1 + e^{-2y}} - \frac{1 - v_Te^{-(y-y_p)}\cos\phi}{1 - 2v_Te^{-(y-y_p)}\cos\phi + e^{-2(y-y_p)} } \biggr \}.
\end{split}
\ee

\subsection{Sum of bulk, bremsstrahlung and interference contributions}
\label{totaldist}

Finally, we calculate numerically the total gluon distribution for the process $q + A \rightarrow X + q +g$. This is obtained either directly computing the main result in \eq\nr{finresspectra} or by summing the bulk, bremsstrahlung and interference contributions in \eqs\nr{bulkspectranum}, \nr{bremssfullv2} and \nr{interf4}:
\be
\label{fulldistrib}
\left ({16\pi^3\over g^2C_F}\right ){{dN} \over {dy d^2k}}\bigg\vert_{\rm total}  \equiv  \left ({16\pi^3\over g^2C_F}\right )\biggl [{{dN} \over {dy d^2k}}\bigg\vert_{\rm bulk}  + {{dN} \over {dy d^2k}}\bigg\vert_{\rm bremss}  + {{dN} \over {dy d^2k}}\bigg\vert_{\rm interf} \biggr ].
\ee
\begin{figure}[t!]
\begin{center}
\includegraphics[width=0.48\textwidth]{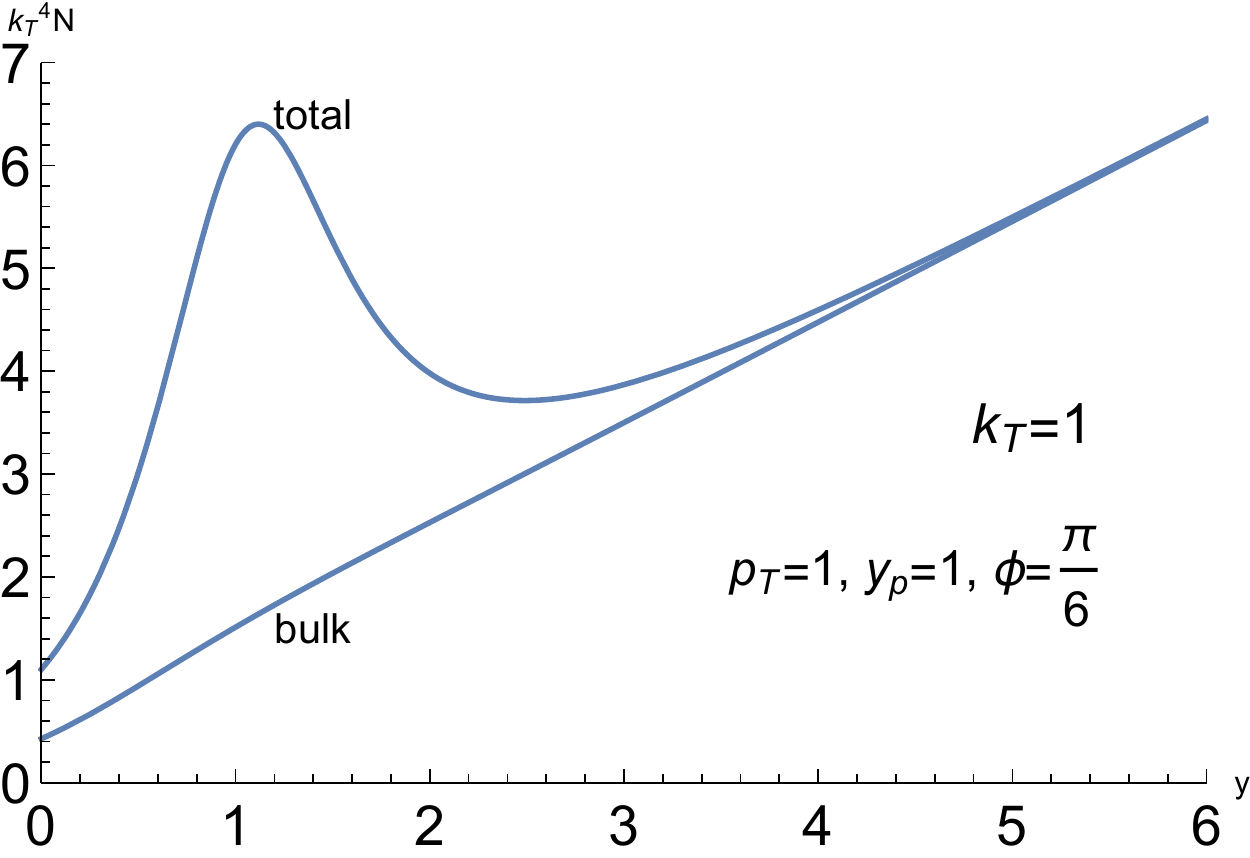}
\includegraphics[width=0.48\textwidth]{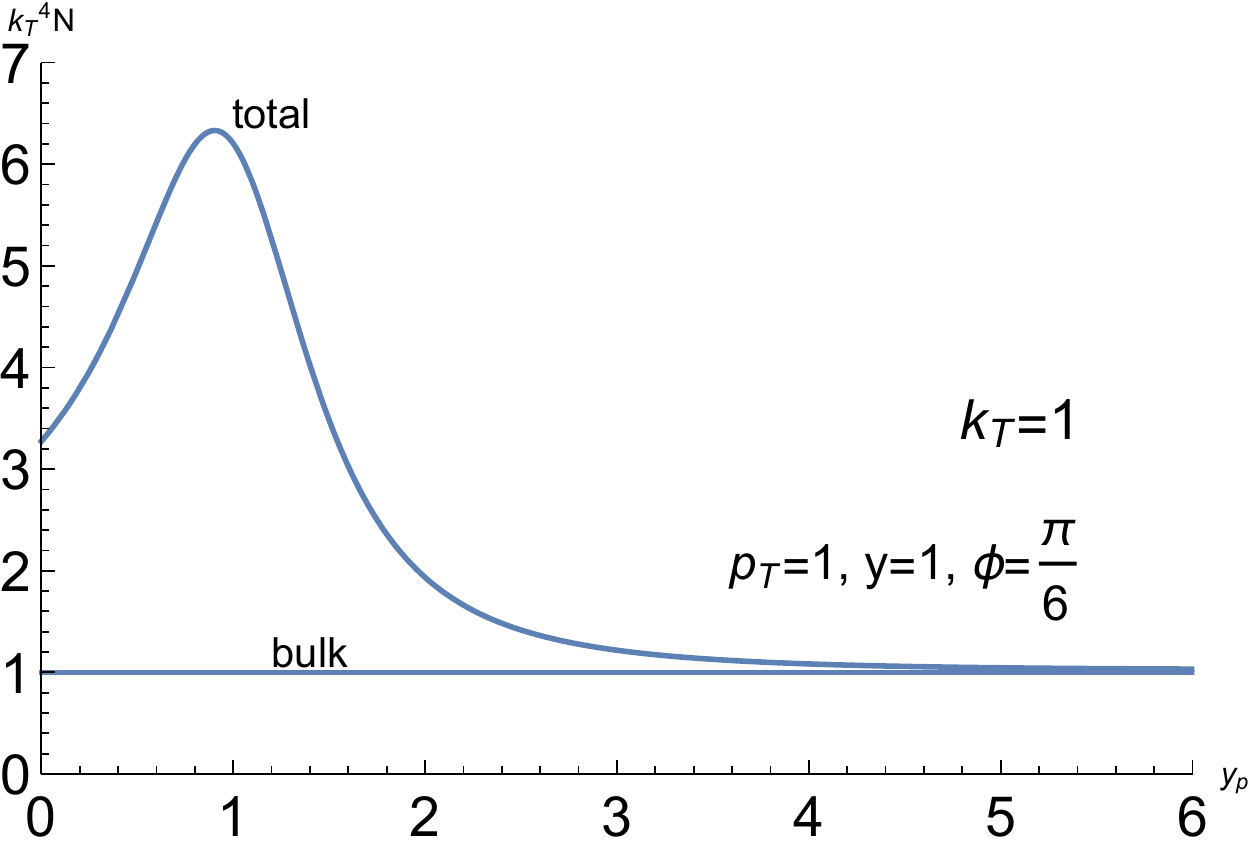}
\end{center}
\caption{\small A plot (left panel) of scaled total gluon distribution \eq\nr{finresspectra} as a function of $y$ for fixed $k_T, p_T, y_p$ and $\phi$. The curve marked bulk is the $k_T=1$ rapidity distribution computed from \nr{bulkspectra} (cf. \fig\ref{yfig}). Right panel is the same distribution but now plotted as a function of $y_p$ for fixed $k_T, p_T, y$ and $\phi$. Here the only reference is that we know that bulk is independent of recoil parameters.}
\label{totaldistfig}
\end{figure}
The total result in \eq\nr{fulldistrib}  can be rewritten in a very compact form
\be
\begin{split}
& \left ({16\pi^3\over g^2C_F}\right ){{dN} \over {dy d^2k}}\bigg\vert_{\rm total} =  2 \int_{0}^{\infty} dh_T \frac{h_T S(h_T)}{2\pi}  \frac{1}{k_T^2}\biggl[ {k_T^2[(h_T^2+k_T^2+a^2)(h_T^2+k_T^2)-4k_T^2h_T^2] \over K^{3/2}}\\
& + \frac{1-v_Te^{\Delta y }\cos\phi}{1 - 2v_T e^{\Delta y}\cos\phi + e^{2\Delta y }}\biggl({h_T^2-k_T^2+a^2\over K^{1/2}}-1\biggr)  +  \frac{1 - 2v_T e^{\Delta y}\cos\phi + v_T^2e^{2\Delta y}}{(1 - 2v_T e^{\Delta y}\cos\phi + e^{2\Delta y})^2}  \biggr],
\end{split}
\ee
where $\Delta y = y_p - y$, $a^2 = k_T^2e^{-2y}$ and the factor $K$ is defined in \eq\nr{factorK}.

In \fig\ref{totaldistfig} (left panel) we show the numerically computed and $k_T$-scaled (though for $k_T=1$ this has no effect)
total gluon distribution as a function of $y$ for fixed values of $k_T, p_T, y_p $ and $\phi$. The right panel shows the same distribution as function of $y_p$ for fixed values of $k_T, p_T, y $ and $\phi$. The quark mass is always fixed to $m =0.2$ and $Q_s^A=2$. In both panels the baseline is given by the bulk curve which should be compared with those in \fig\ref{yfig}: at this small $k_T$ one does not observe any turnover to the central plateau. On top of the bulk curve one observes a strong peak structure from bremsstrahlung and its interference with bulk.
The detailed form of the peak is sensitively dependent on the relative magnitudes of $\kt$ and the recoil momentum $\pt$. If $\phi=0$, one can even
hit the $p\cdot k=0$ singularity at $y=y_p$. In the figure $\phi=\pi/6$ and increasing it further the peak structure gets less dominant. This peak structure is
a definite experimental prediction of the model, which, of course, is modified by quantum mechanics and the fact that the quark in practice
has to be embedded in a hadron.

%% file: conclu.tex
\section{Conclusions}
We have in this paper studied QCD dynamics in the usual CGC formalism but 
in the rather unusual setting of a very large energy nucleus
colliding with a static classical colored particle, called a quark. The collision kicks the initial quark to some constant
final momentum $p^\mu$.
Physically one has in mind a study of nucleus-nucleus collisions in the rest frame of
either of the nuclei, i.e., in the target or beam fragmentation regions. The rationale is that it might be useful to
dissect the problem in microscopic pieces and the extreme clearly is to just have a single quark as a target.
From this one can then proceed to the level of nucleons and finally nuclei.

The outcome is a rather compact formula describing the final gluonic radiation field, from which the distribution
of gluons is computed by complex squaring and averaging over an ensemble of sources. The result can be split in
two components. The first, the bulk, arises mainly from the interaction of the nuclear color field with the color electric 
field of the nucleus, is independent of the momentum of the kicked quark and gives the gluon $k_T$ and 
longitudinal rapidity $y$ distribution in the range we are interested in. At fixed $k_T$ 
the $y$ distribution grows monotonically 
from negative values of $y$ up to $y\approx \log(k_T/m_\rmi{quark})$ where it levels off to a central region plateau.
The second component, bremsstrahlung, is in the classical approximation
independent of interactions with the nucleus, depends on the kicked quark momentum $p^\mu$ and decreases too
slowly with increasing $k_T$. At very large $k_T$ it has to be improved by including recoil effect in the scattered quark.
However, at smaller $k_T\ll Q_s^A$, where most of the gluon production takes place, it should give a good 
description of physics. We find that at rapidities in the target framentation region it gives rise to a significant peak,
somewhat dependent on the recoil momentum.
At large $y$ bremsstrahlung decays $\sim e^{-2y}$ and thus the bulk contribution dominates in the
central region.

The outcome should now be usable to estimate energy deposition from gluon radiation in the target fragmentation region,
as input to initial conditions of hydrodynamical evolution of QCD matter formed. Baryon number initial values
could come from the Dirac equation solutions in \cite{McLerran:2018avb}.
However, the inclusion of recoil effects in the classical approximation is still incomplete. One obvious avenue for
progress is computing joint production rates of gluons and quarks by extending the computation in \cite{marquet}
to the target fragmentation region.

%% file: appA.tex
\begin{figure}[!b]
\begin{center}

\includegraphics[width=0.48\textwidth]{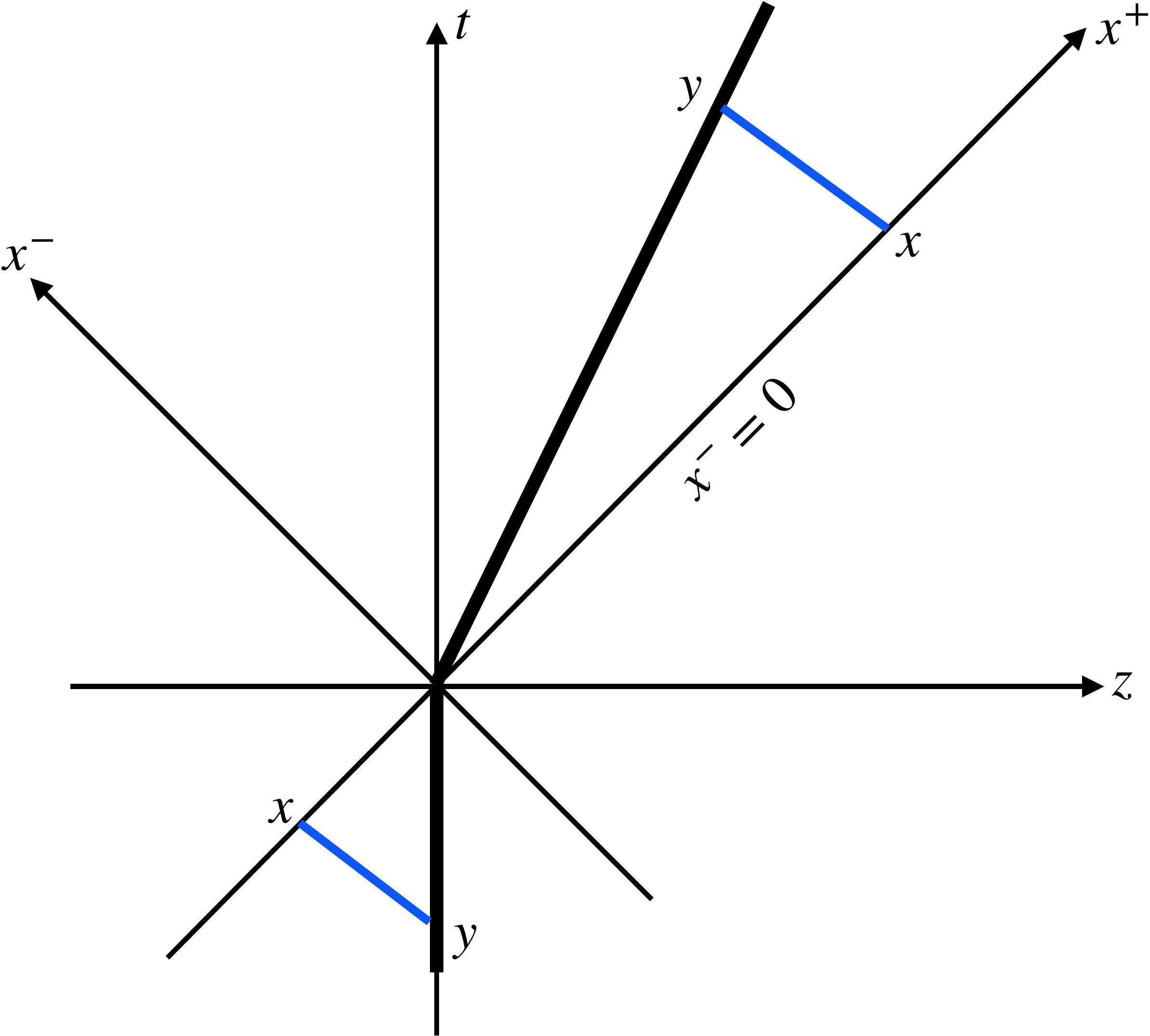}

\end{center}

\caption{\small Computation of the two pieces \nr{aifinapp} and \nr{aifininiapp}
of the radiation field along $x^-=0$. The first piece is
sourced by the kicked quark and brought to $x^-=0$ by the advanced Green's function and the second is 
sourced by the initial quark and brought to $x^-=0$
by the retarded Green's function.
}
\label{bremsint}
\end{figure}

In this section, we derive the results used in \eq\nr{vu}. Starting from the most difficult case, the task is to compute the advanced 
Green's function integral towards the line $x^-=0$ over the scattered quark
(see Fig. \ref{bremsint})
\be
\begin{split}
\label{aifinapp}
a^i_{\rm fin}(x^+, x^-=0,\xt) & \equiv \int d^4y~G_A(x-y)\mathcal{J}^{i}_{\rm eff}(y^+,y^->0,\yt).
\end{split}
\ee
From the derivative of this with respect to $x^+$ its Fourier transform over the $x^+$ can be computed
\be
\label{ufuncapp}
\begin{split}
u^i(\xt) \equiv 2i\int dx^+  e^{ik^-x^+} \partial_{+}a^{i}_{\rm fin}(x^+,x^-=0,\xt),
\end{split}
\ee
where the advanced Green's function in the LC coordinates is given by 
\be
G_A(x-y) = +\frac{1}{2\pi}\theta(x^- - y^-)\delta((x-y)^2)
\ee
and the source term $\mathcal{J}^{i}_{\rm eff}(y^- > 0)$ (defined in \eq\nr{jiieff}) corresponds to the kicked quark current \eq\nr{bigj2}, 
\be
\label{sourcekickapp}
\mathcal{J}^{i}_{\rm eff}(y^+,y^->0,\yt)  = \left (\frac{p^i}{p^-} + \partial^i \frac{1}{\partial_+}\right )\delta\left (y^+ - \frac{p^+}{p^-}y^-\right )\delta^{(2)}\left (\yt - \frac{\pt}{p^-}y^- \right ).
\ee
Since the source term in \eq\nr{sourcekickapp} contains the inverse of the operator $\partial_+$, it is actually simpler to solve for $\partial_+ a^i_{\rm fin}$. Also the Fourier transform of 
 $\partial_+ a^i_{\rm fin}$ in the $x^+$ direction converges, not that of $a^i_{\rm fin}$ as such. Inserting these explicitly into \eq\nr{aifinapp} yields  
 \be
 \begin{split}
2\pi \partial_+ a^i_{\rm fin}(x^+,x^-,\xt) = \int dy^+dy^-d^2y\,\theta(y^--x^-)\delta((x-y)^2)\left (\fra{p_i}{p^-}\partial_+^y+\partial_i^y \right )\\
\times \delta\left (y^+ - \frac{p^+}{p^-}y^-\right )\delta^{(2)}\left (\yt - \frac{\pt}{p^-}y^- \right ).
 \end{split}
 \ee

The derivative of the delta functions can be dealt with by partial integration by moving the derivative bracket two steps to the left (with sign change)
and then converting $\partial^y$ to $-\partial^x$ (since integrand is a function of $y-x$) and taking it outside the integral. Using the
other delta functions the light cone delta function  becomes a quadratic polynomial in $y^+$:
\be
\delta((x-y)^2) = \delta \left ({p^2\over (p^+)^2}(y^+-y^+_+)(y^+-y^+_-)\right ),
y^+_\pm =-{p^+\over m^2}\left (p\cdot x\pm\sqrt{(p\cdot x)^2+m^2x^2} \right ).
\ee
The scalar products here are 4-dimensional. On the interesting line $x^-=0$ we have $x^2=x_T^2$, the two roots have opposite
signs so that only one of them contributes (\fig\ref{bremsint} shows that $y^+>0$). Collecting all factors the result is
\ba
\partial_+a^i_{\rm fin}(x^+,0,\xt)&=&{1 \over 4\pi}(p_i\partial_++p^-\partial_i){1\over \sqrt{(-p^-x^++\pt\cdot\xt)^2+m^2x_T^2}}\nn
&=&-{1 \over 4\pi}m^2p^-{x^i\over[(-p^-x^++\pt\cdot\xt)^2+m^2x_T^2]^{3/2}}.
\label{GAsol}
\ea
Substituting this into \eq\nr{ufuncapp} we find 
\be
\begin{split}
u^i(\xt) = -\frac{ix^i}{2\pi} m^2p^-\int dx^+ \frac{e^{ik^-x^+}}{[(p^-)^2(x^+ - \frac{1}{p^-}\pt \cdot \xt)^2+m^2x_T^2]^{3/2}},
\end{split}
\ee
where the integrand is written in a form which suggests replacing $x^+$ by
\be
z^+ = x^+-\fra1{p^-}\pt\cdot\xt.
\ee
This also creates a transverse momentum shift $k_i\to k_i-\fra{k^-}{p^-}p_i$ in the $\xt$ integral. Integration over
$z^+$ is straightforward and gives the result
\be
u^i(\xt) = -\frac{i}{2\pi} \frac{2mk^-}{p^-}\frac{x^i}{x_T}\exp \biggl [i\frac{k^-}{p^-}\pt \cdot \kt \biggr ] K_1\left (\frac{mk^-}{p^-}x_T \right ). 
\ee

Furthermore, we also need the Fourier transform of $u^i(\xt) \to u^i(\kt)$.  The integration over the two-dimensional transverse space can be easily performed by first noting that
\be
\int d^2x e^{-i\kt \cdot \xt } x^i f(\xt) = -2i\pi \frac{k^i}{k_T}\int_{0}^{\infty} dx_T x_T^2 f(\xt) J_1(k_Tx_T),
\ee
where $J_1$ is the modified Bessel function of the second kind. This yields the expression 
\be
\begin{split}
\int d^2x e^{-i\kt \cdot \xt } u^i(\xt) & = -\frac{2mk^-}{p^-}\frac{r^i}{r_T}\int_{0}^{\infty}d x_T x_T K_1\left (\frac{mk^-}{p^-}x_T \right ) J_1\left (r_Tx_T\right ),
\end{split}
\ee
where we have defined $r^i = k^i - (k^-/p^-)p^i$. Using this result and performing the remaining integral we find
\be
u^i(\kt) = { -2(k^i-\fra{k^-}{p^-}p^i)\over | k^i-\fra{k^-}{p^-}p^i|^2+(\fra{k^-}{p^-})^2m^2}.
\label{uki}
\ee

Let us then consider the easier case, the retarded Green's function integral towards the line $x^-=0$ over the initial static quark
(see Fig. \ref{bremsint})
\be
\begin{split}
\label{aifininiapp}
a^i_{\rm ini}(x^+, x^-=0,\xt) & \equiv \int d^4y~G_R(x-y)\mathcal{J}^{i}_{\rm eff}(y^+,y^- < 0,\yt).
\end{split}
\ee
From this one can then compute the other function in \eq\nr{vu}:
\be
\label{vfuncapp}
\begin{split}
v^i(\xt) \equiv 2i\int dx^+  e^{ik^-x^+} \partial_{+}a^{i}_{\rm ini}(x^+,x^-=0,\xt),
\end{split}
\ee
where the retarded Green's function is given by 
\be
G_R(x-y) = +\frac{1}{2\pi}\theta(y^- - x^-)\delta((x-y)^2).
\ee
The source term $\mathcal{J}^{i}_{\rm eff}(y^- <0)$ corresponding to the initial static quark reads
\be
\mathcal{J}^{i}_{\rm eff}(y^+,y^-< 0,\yt)  =  \partial^i \frac{1}{\partial_+} \delta\left (y^+ -  y^-\right )\delta^{(2)}\left (\yt  \right ).
\ee
Following the same steps as before we find 
\be
v^i(\xt) = -\frac{i}{2\pi} 2\sqrt 2 k^- \frac{x^i}{x_T} K_1\left (\sqrt 2 k^- x_T \right ). 
\ee
Fourier transforming this to momentum space gives (this follows from \eq\nr{uki} with $p_i=0$ and $p^-=m/\sqrt2$)
\be
\label{vki}
v^i(\kt) = { -2k^i \over k_T^2 + 2(k^-)^2}.
\ee

%% file: appC.tex
There is extensive literature on the calculation of the correlation function of Wilson lines, see for example \cite{gelispeshier,blaizotgv,Gelis:2018qru}.  In this section, we outline how to compute average over the color sources in the 
correlation function $S(\xt-\yt)$ of two Wilson lines introduced in \eq\nr{wcorrelators}. All of this is well known material, but the discussion in
this paper relies so heavily on it that maybe it is useful to summarize its derivation.

To calculate gauge invariant observables $\mathcal{O}$ in terms of the classical color field we must perform average over the set of sources $\rho$, defined as
\be
\label{eq:Oaverho}
\langle \mathcal{O} \rangle_\rho \equiv \int \mathcal{D}\rho ~W [\rho] \mathcal{O} [\rho],
\ee
where the weight function $W [\rho]$ is the distribution of color sources. In this work, we use the simple MV model approximation in which $W[\rho]$ is obtained by considering a Gaussian distribution   
\begin{equation}
W[\rho(\xt)] = \exp \biggl [-\int d^2 x \frac{\rho^2(\xt)}{2\lambda} \biggr ],
\label{rhodist}
\end{equation}
where $\lambda$ describes the color source density of the nucleus and we assume that the color sources only depends on the transverse position $\xt$.

First, let us consider the 2-point correlation function  
\be
\label{Sdefapp}
S(\xt-\yt) \equiv \frac{1}{N^2-1} 	\bigg\langle \tr(\uw(\xt)\uw^{\dagger}(\yt))\bigg\rangle_{\rho},
\ee
where the adjoint Wilson line $\uw(\xt)$ in the eikonal approximation simplifies to 
\be
\label{Wilsonapp}
\uw(\xt) = \exp \biggl [-igT^a_{A}\Phi_a(\xt)\biggr ]
\ee
with $\Phi(\xt) = \frac{1}{\partial^2_T} \rho_a(\xt)$. Using the relation 
\begin{equation}
\frac{1}{\partial^2_T}\rho_a(\xt) = \int d^2 z\, G(\xt-\zt) \rho_a(\zt),
\end{equation}
where $G$ is the Green's function associated with the 2-dimensional Laplacian and satisfies 
\begin{equation}
G(\xt-\zt) = \int \frac{d^2 k}{(2\pi)^2} \frac{e^{i\kt\cdot (\xt-\zt)}}{k_T^2},
\label{2dgreen}
\end{equation}
allows us to rewrite the Wilson line in \eq\nr{Wilsonapp} as 
\begin{equation}
\uw(\xt) = \exp \biggl [-igT^a_A \int d^2 z\, G(\xt-\zt)\rho_a(\zt) \biggr ].
\end{equation}
Now the expectation value of two adjoint Wilson line in \eq\nr{Sdefapp} can be written as 
\begin{eqnarray}
\bigg\langle \uw(\xt) \uw^\dagger(\yt) \bigg\rangle_\rho  = \int  \mathcal{D}\rho \exp \biggl [-\int d^2 z \biggl (\frac{1}{2\lambda}\rho^2+ ig \left (G(\xt-\zt)-G(\yt-\zt)\right )\rho \biggr )\biggr ].
\label{correv1}
\end{eqnarray}
Performing the Gaussian integral leads to the result
\begin{eqnarray}
\bigg\langle \uw(\xt) \uw^\dagger(\yt)  \bigg\rangle_\rho  =  \exp \biggl [-\frac{\lambda g^2}{2}(T^a_AT^a_A)\int d^2 z \biggl (G(\xt-\zt)-G(\yt-\zt) \biggr )^2\biggr ],
\label{correv2}
\end{eqnarray}
where the matrix  $T^a_AT^a_A = C_A \mathbf{1}_{N^2-1}$ is the adjoint Casimir $C_A=N$ times an adjoint unit matrix $\mathbf{1}_{N^2-1}$. Exponentiation 
is trivial and we have
\be
S(\xt-\yt)={1\over N_c^2-1}\bigg\langle \tr (\uw(\xt) \uw^\dagger(\yt) )  \bigg\rangle_\rho=\exp \biggl [-Q_s^2 D(\vert\xt-\yt\vert)\biggr ],
\ee
where $Q_s^2=\fra12\lambda g^2 C_A$ and by using the integral representation of the Green's function we have defined the dipole function $D$ as
\be
D(|\xt-\yt|) \equiv
\int d^2 z \, \biggl (G(\xt-\zt) - G(\yt-\zt) \biggr )^2 =  \int \frac{d^2k}{(2\pi)^2}\, \frac{2}{(k_T^2)^2}\biggl [1 - e^{i\kt\cdot (\xt-\yt)} \biggr ].
\label{D}
\ee
The remaining Fourier integral in \eq\nr{D} is IR divergent. This can for example be regulated by replacing $k_T^2$ by $k_T^2 + m_{\rm IR}^2$:
\begin{equation}
D_m(\rt)=  \int \frac{d^2k}{(2\pi)^2}\, \frac{2}{(k_T^2 + m_{\rm IR}^2)^2}\biggl (1 - e^{i\kt\cdot\rt} \biggr )=
{1\over 2\pi m_{\rm IR}^2}\biggl[1-m_{\rm IR}r_TK_1(m_{\rm IR}r_T)\biggr],
\label{reg}
\end{equation}
where $\rt = \xt-\yt$ and $r_T = \vert \xt-\yt\vert$.

Similarly, for the expectation value of a single adjoint Wilson line one obtains
\ba
\bigg\langle \uw(\xt) \bigg\rangle_{\rho}&=&\exp \biggl [-Q_s^2\, \mathbf{1}_{N^2-1}\int d^2zG^2(\xt - \zt) \biggr ]\nn
&=& \mathbf{1}_{N^2-1}\exp\left[-Q_s^2\int {d^2x\over 16\pi^2}\log^2\left({1\over x^2\Lambda^2}+1\right)\right]\nn
&=& \mathbf{1}_{N^2-1}\,\exp[-{\pi\,Q_s^2\over 48 \Lambda^2}],
\ea
where we regulated IR and UV of the Green's function \nr{2dgreen} by
\be
G(\xt)={1\over 4\pi}\log\left({1\over x_T^2\Lambda^2}+1\right).
\ee
Compating with mass-regulated Green's function, $\Lambda=\fra12 m_{\rm IR} e^{\gamma_E}$. The expectation value of a 
single Wilson line thus effectively vanishes.

\subsection*{Asymptotic behaviour}

\begin{figure}[!t]
\begin{center}
\includegraphics[width=0.6\textwidth]{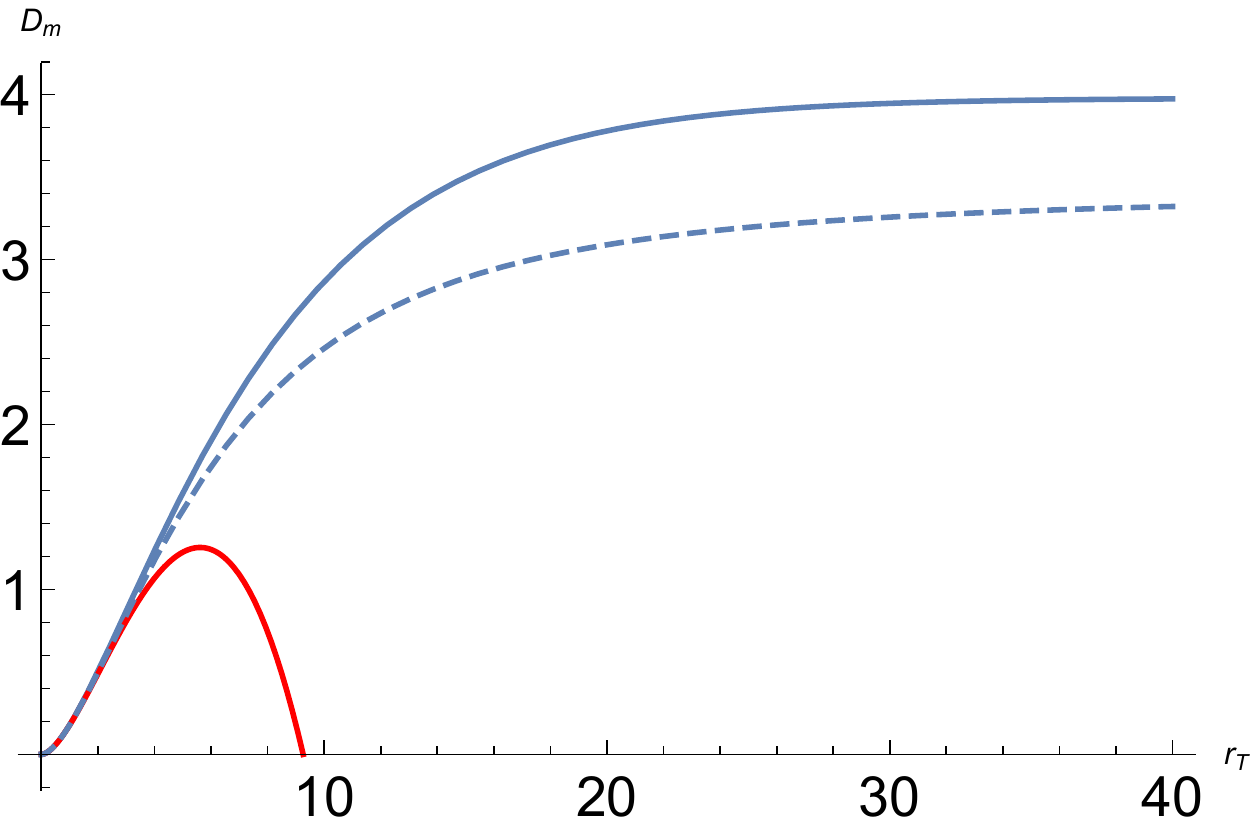}
\end{center}
\caption{\small The function $D_m(r_T)$ in \eg\nr{reg} and its small $m_{\rm IR}$ approximation in \eq\nr{regsmall} plotted for $m_{\rm IR}=0.2$. The maximum of the
red curve is at $r_T=e^{-1/2}/\Lambda_{\rm QCD}$. The dashed curve is the approximation \nr{DLa}.
}
\label{massreg}
\end{figure}

Let us then study the asymptotic behaviour of \eq\nr{reg}. Expanding in small $m_{\rm IR}^2$ gives 
\begin{eqnarray}
D_{m}(r_T) &=& \frac{r_T^2}{4\pi} \biggl (\log \left (\frac{2}{r_Tm_{\rm IR}}\right ) -\gamma_E + \frac{1}{2} \biggr  ) + \mathcal{O}(m_{\rm IR}^4)
\equiv  \frac{r_T^2}{4\pi} \log \left (\frac{1}{r_T\Lambda_{\rm QCD}}\right ),
\label{regsmall}
\end{eqnarray}
where $\Lambda_{\rm QCD}=\fra12 e^{\gamma_E-\fra12}m_{\rm IR}=0.5401\,m_{\rm IR}$. How the mass regularisation operates is clearly seen from \fig\ref{massreg}, where the function \eq\nr{reg} and its small $m_{\rm IR}$
approximation \nr{regsmall} are plotted for $m_{\rm IR}=0.2$. The maximum of the small $m_{\rm IR}$ approximation is at $r_T=2e^{-\gamma_E}/m_{\rm IR}=1.123/m_{\rm IR}$. \fig\ref{massreg} also shows that another simple way of extending the small $r_T$ approximation \nr{regsmall} to larger $r_T$ is by replacing it by
\be
D_\Lambda(r_T)=  \frac{r_T^2}{8\pi} \log \left (\frac{1}{(r_T\Lambda)^2}+1\right ).
\label{DLa}
\ee

\begin{figure}[!t]
\begin{center}
\includegraphics[width=0.48\textwidth]{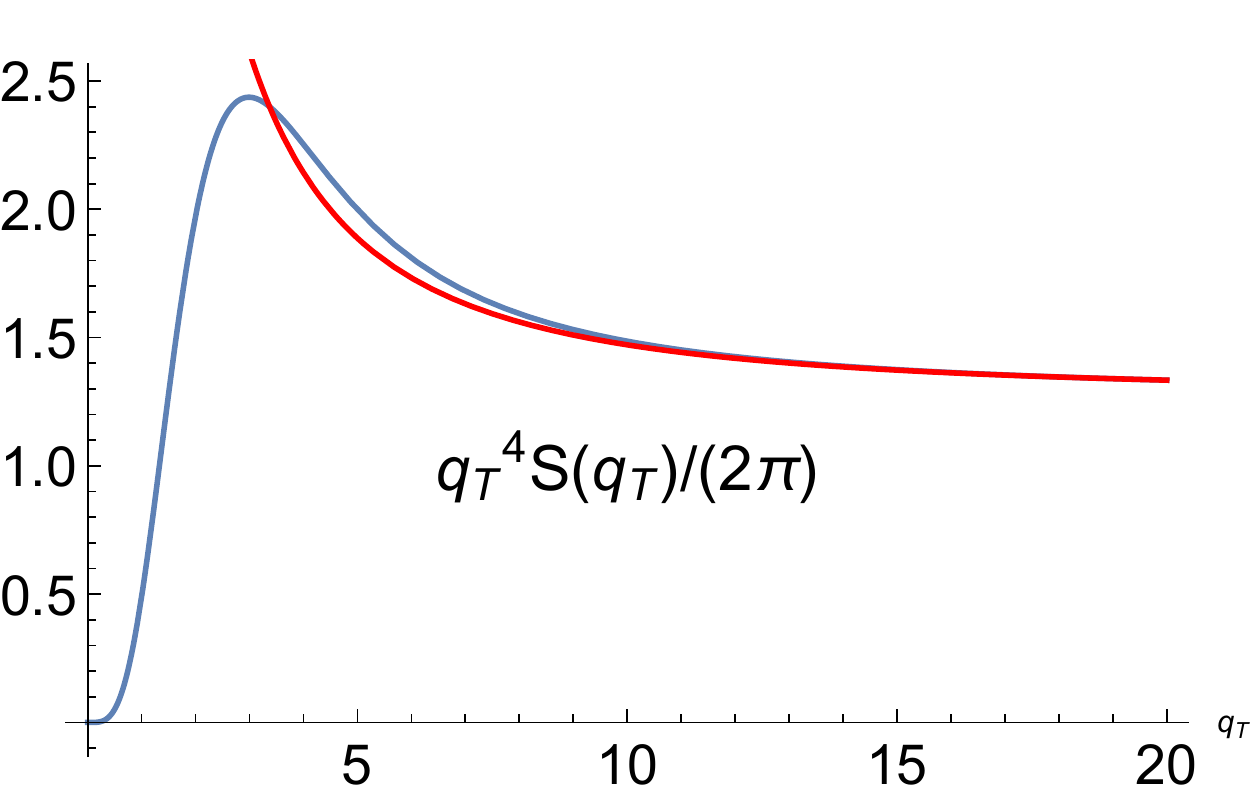}
\includegraphics[width=0.48\textwidth]{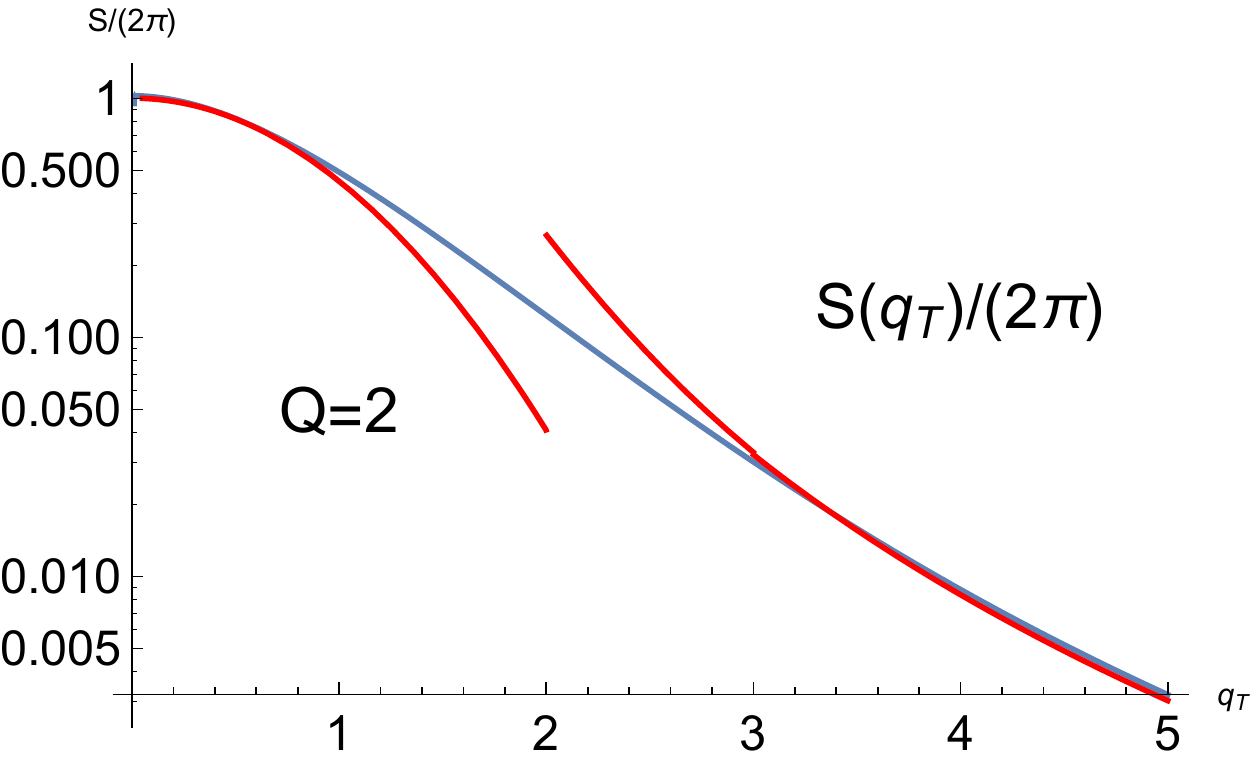}
\end{center}
\caption{\small The left panel shows plot of  $q_T^4S(q_T)/(2\pi)$ for $Q^A_s=2,\,\Lambda=0.10802$ 
together with the large $q_T$ approximation \nr{gp} with the logarithmic factor $\log(q_T/0.468)$ (red).
The right panel shows unscaled $S(q_T)$ together with a gaussian  $q_T<2$ approximation $1.025\exp(-0.8q_T^2)$
and the large $q_T>2$ approximation.
}
\label{SqTfig}
\end{figure}

For the Fourier transform of correlation function $S(r_T)$ we have
\ba
S(q_T)&=&\int d^2r\,e^{-i\qt\cdot\rt}S(r_T)=2\pi \int_0^\infty dr\,rJ_0(q_Tr)\exp[-Q^2D(r)]\\
&=&{2\pi\over q_T^2} \int_0^\infty dr\, r J_0(r)\exp\left[-{Q^2\over 8\pi q_T^2}r^2\log\left({q_T^2\over (0.10802\, r)^2}+1\right)\right],
\label{Sfourier}
\ea
where in the second form $\Lambda$ regularisation \eq\nr{DLa} with $\Lambda=0.5401m_{\rm IR}=0.10802$ corresponding to $m_{\rm IR}=0.2$
has been used. Because of the Bessel function the integral is rapidly varying and special methods can be developed for
accurate evaluation \cite{hankel}. In Mathematica NIntegrate with Method $\to$ "ExtrapolatingOscillatory" can be applied.
At large $r$ the integrand is exponentially small, $\sim \exp(-Q^2/(2\pi m_{\rm IR}^2))$. 

For $q_T\gg Q^A_s$, the correlation function $S(q_T)$ goes like \cite{gelispeshier},
\be
S(q_T)={2(Q^A_s)^2\over q_T^4}+{8(Q^A_s)^4\over \pi\,q_T^6}\left(\log{q_T\over\Lambda_{\rm QCD}}-1\right).
\label{gp}
\ee
This can be derived by expanding the exponent in \eq\nr{Sfourier} with mass regularisation:
\be
\begin{split}
S(q_T) &=\int d^2r\,e^{-i\qt\cdot\rt}\biggl [1-(Q^A_s)^2\int {d^2p\over(2\pi)^2}{2\over (p_T^2+m_{\rm IR}^2)^2}(1-e^{i\pt\cdot\rt})\\
& + \frac{(Q^A_s)^4}{2}\int{d^2p_1\over (2\pi)^2}{d^2p_2\over (2\pi)^2}{4\left (1-e^{i{\bf p_1}\cdot\rt}-e^{i{\bf p_2}\cdot\rt}+e^{i({\bf p_1}+{\bf p_2})\cdot\rt} \right )\over (p_{1T}^2+m_{\rm IR}^2)^2(p_{2T}^2+m_{\rm IR}^2)^2}
\biggr ] \\
& ={2(Q^A_s)^2\over q_T^4} + \frac{(Q^A_s)^4}{2}\left[-{2\over \pi m_{\rm IR}^2q_T^4}+4\int{d^2p\over (2\pi)^2}\,
{1\over (p^2+m_{\rm IR}^2)^2( |\qt-\pt|^2+m_{\rm IR}^2)^2}\right],
\end{split}
\ee
where the $\rt$ integral is carried out first  and in the remaining $\pt$ integral one has taken the limit $m_{\rm IR}\to 0$; the powerlike $1/m_{\rm IR}^2$ divergence then cancels. This gives the result in \eq\nr{gp}, where the scale of the log depends on the regularisation method used.

\fig\ref{SqTfig} (left panel) shows a plot of $q_T^4S(q_T)/(2\pi)$ for $Q^A_s=2$ and $\Lambda=0.10802$
together with the approximation \nr{gp} combined with the logarithmic factor $\log(q_T/0.468)$. This scale of the log gives a very good fit in the large $q_T$ range but undershoots below. This is an effect of order $Q^6/q_T^8$. 
The right panel shows $S(q_T)/(2\pi)$
without the scaling together with a small $q_T$ fit $1.025\exp(-0.8q_T^2)$ and the large $q_T$ approximation \nr{gp}. 
The transition region is at $q_T\approx Q^A_s=2$.

%% file: appB.tex
Here we show how to calculate color average over the classical color charges in \eq\nr{eq:radaverho}.  After performing the square in \eq\nr{eq:radaverho}, it is clear that there are four different color structures to be evaluated:
\be
\begin{split}
&\bigg\langle  \uw_{ad}(\xt)\uw_{db}^{\dagger}(\ot)Q^bQ^c\uw_{ce}(\ot)\uw_{ea}^{\dagger}(\yt)  \bigg\rangle_Q,\\
&\bigg\langle  \uw_{ad}(\xt)\uw_{db}^{\dagger}(\ot)Q^bQ^a  \bigg\rangle_Q,\\
&\bigg\langle  Q^aQ^c\uw_{ce}(\ot)\uw_{ea}^{\dagger}(\yt) \bigg \rangle_Q,\\
&\bigg\langle  Q^aQ^a \bigg\rangle_Q.
\end{split}
\ee
By using the relation given in \eq\nr{coloraverage} and the unitarity condition between two adjoint Wilson lines $\uw(\ot)\uw^{\dagger}(\ot) = \mathbf{1}_{N^2-1}$ we obtain the results 
\be
\label{coloralgebraapp}
\begin{split}
\bigg\langle  \uw_{ad}(\xt)\uw_{db}^{\dagger}(\ot)Q^bQ^c\uw_{ce}(\ot)\uw_{ea}^{\dagger}(\yt)  \bigg\rangle_Q &= \frac{1}{2} \tr (\uw(\xt)\uw^{\dagger}(\yt)),\\
\bigg\langle  \uw_{ad}(\xt)\uw_{db}^{\dagger}(\ot)Q^bQ^a  \bigg\rangle_Q &= \frac{1}{2} \tr (\uw(\xt)\uw^{\dagger}(\ot)),\\
\bigg\langle  Q^aQ^c\uw_{ce}(\ot)\uw_{ea}^{\dagger}(\yt) \bigg \rangle_Q &= \frac{1}{2} \tr (\uw(\yt)\uw^{\dagger}(\ot)),\\
\end{split}
\ee
and $\bigg\langle  Q^aQ^a \bigg\rangle_Q = \frac{N^2-1	}{2}$. By substituting these results into \eq\nr{eq:radaverho} gives the expression in \eq\nr{intres}.